# The effects of HIV self-testing on HIV incidence and awareness of status among men who have sex with men in the United States: Insights from a novel compartmental model

*Alex Viguerie, PhD[a,*]*

*Chaitra Gopalappa, PhD[a,b]*

*Cynthia M. Lyles PhD[a]*

*Paul G. Farnham, PhD[a]*

[a]Centers for Disease Control and Prevention (CDC), National Center for HIV, Viral Hepatitis, STD, and TB Prevention (NCHHSTP), Division of HIV Prevention (DHP), Quantitative Sciences Branch (QSB)
[b]Department of Mechanical and Industrial Engineering, University of Massachusetts Amherst, Amherst, MA 01003 (USA)
[*]Corresponding author: xjm9@cdc.gov , +1 (404) 718-2691

# The effects of HIV self-testing on incidence and awareness of HIV status among men who have sex with men in the United States: Insights from a novel compartmental model

**Abstract**

The OraQuick In-Home HIV self-test represents a fast, inexpensive, and convenient method for users to assess their HIV status. If integrated thoughtfully into existing testing practices, accompanied by efficient pathways to formal diagnosis, self-testing could both enhance HIV awareness and reduce HIV incidence. However, currently available self-tests are less sensitive, particularly for recent infection, than gold-standard laboratory tests. It is important to understand the impact if some portion of standard testing is replaced by self-tests. We introduced a novel compartmental model to evaluate the effects of self-testing among gay, bisexual and other men who have sex with men (MSM) in the United States for the period 2020 to 2030. We varied the model for different screening rates, self-test proportions, and delays to diagnosis for those identified through self-tests to determine the potential impact on HIV incidence and awareness of status.

When HIV self-tests are strictly supplemental, self-testing can decrease HIV incidence among MSM in the US by up to 10% and increase awareness of status among MSM from 85% to 91% over a 10-year period, provided linkage to care and formal diagnosis occur promptly following a positive self-test (90 days or less). As self-tests replace a higher percentage laboratory-based testing algorithms, increases in overall testing rates were necessary to ensure reductions in HIV incidence. However, such increases were small (under 10% for prompt engagement in care and moderate levels of replacement). Improvements in self-test sensitivity and/or decreases in the detection period may further reduce any necessary increases in overall testing. Our study suggests that, if properly utilized, self-testing can provide significant long-term reductions to HIV incidence and improve awareness of HIV status.

**Introduction**

Since the FDA approved the OraQuick In-Home HIV test in 2012, the use of HIV self-testing has grown considerably in the United States, amplified in part by the COVID-19 pandemic and reliance on telehealth and telemedicine services, such as COVID-19 home tests [1]. In comparison to non-FDA approved HIV self-collection kits that collect a dried blood spot (DBS) sample, oral swab self-testing gives test results within minutes and does not require a sample to be sent to an external laboratory [2]. HIV self-testing offers an innovative strategy to help expand HIV testing across the United States, increase testing among populations most affected by HIV, and reach persons with undiagnosed HIV who are not being reached by traditional testing programs, including those in laboratory settings [3], [4].There is currently only one FDA-approved HIV self-test in the US [5], [6].

Several studies have demonstrated the effectiveness of self-testing programs. In [4], a trial among gay, bisexual and other men who have sex with men (collectively referred to as MSM) showed that distribution of oral self-tests led to a 47% increase in total HIV testing over the observed period. The availability of self-testing was not found to have any impact on STI acquisition and performed similarly to laboratory-based testing for all HIV risk indicators. A trial utilizing internet-based recruitment, in which participants were mailed self-tests [7], found that this approach led to increases in testing levels especially among MSM who had never had a previous HIV test [3]. This is generally consistent with existing literature, which suggests that overall testing increases with improved availability of self-testing [8], [9]. While the cost of the current FDA-approved self-test has been reported to be a deterrent from further adoption of self-testing in the United States [10], [11], recent analyses have found that self-testing is cost saving [12].

Despite encouraging findings in trials, concerns about the real-world impact of scaling-up self-testing programs remain. The FDA-approved self-tests (like all oral-swab tests including CLIA waived rapid tests) are less sensitive than laboratory-based tests, particularly in the acute phase of HIV infection, estimated from 45 to up to 90 days after acquiring infection and referred to as the *detection period* (or *window period*) [5]. While the instructions for the OraQuick self-test state that the test should not be used as a substitute for laboratory-based testing, some researchers have posited a concern that this could happen in practice. In addition, in [4], [7], [8], participants in controlled settings were carefully instructed on the proper use and interpretation of such tests. However, fidelity to proper use may vary with larger-scale implementation in community settings.

Given the inherent difficulty in measuring self-testing behaviors outside of a controlled setting and evaluating the effectiveness of scaling up self-testing programs, researchers have resorted to mathematical models to evaluate the effects of self-testing on the wider population [13], [14]. Findings in these two studies have been mixed, with self-tests having a positive or negative net-effect on HIV incidence based on the degree to which they replace laboratory-based tests and how much their introduction causes overall testing rates to increase. Both studies employed area-specific, network-based models of MSM. In some instances, these studies considered demographic pools that already had high rates of laboratory HIV testing, limiting the potential benefit of any additional testing-based intervention, including self-testing.

Our analysis expanded the mathematical modeling of self-testing by precisely quantifying the relationship between increased HIV self-testing and future HIV incidence and awareness of HIV status in any PWH population. This was accomplished through the introduction of a streamlined mathematical model, that allowed us to vary levels of HIV self-testing, levels of overall HIV testing, and any delays in obtaining a formal HIV diagnosis after a positive self-test, in a straightforward manner. We then applied the model to the US MSM PWH population and identified the conditions in which self-testing could reduce HIV incidence and increase awareness of HIV status. We found *threshold levels,* the amount by which overall HIV testing must increase to offset the reduced sensitivity and delays in obtaining a formal diagnosis associated with increased HIV self-testing. While previous models identified scenarios in which self-testing both reduced and increased HIV incidence, they did not identify the specific conditions in which self-testing leads to HIV incidence reductions.

**Methods**

*Compartmental Model*

Consider a four-compartmental model (Figure 1) where persons with HIV (PWH) are in one of the following states: acute infection unaware of infection ($a$), chronic infection unaware of infection ($u$), AIDS patients unaware of infection ($s$), and all persons with diagnosed HIV infection ($d$). The model is defined here for any PWH population and then applied to the U.S. MSM PWH population in the following section.

This model stratification assumes the following disease progression transitions.

Every person begins in the acute unaware stage $a$, and has one of the following outcomes:

1. The individual can take a test (either self-tests or non-self-tests) and receive a diagnosis, moving to compartment $d$. The time to diagnosis is different for positive self-tests versus non-self-tests in general;
2. The individual can remain unaware and pass from the acute stage to the chronic, unaware stage $u$;
3. The individual may die.

Assuming that (2) happens, the unaware, non-acute individual has one of the following outcomes:

1. The individual can take a test (which may or may not be a self-test), receive a diagnosis, and move to compartment $d$. Again, the time to diagnosis is different for self-tests and non-self-tests;
2. The individual can remain unaware and pass from the chronic unaware stage and develop AIDS, moving to the stage $s$;
3. The individual may die.

Again, assuming (2), an individual sick with AIDS and unaware of their infection:

1. may take a test (possibly a self-test) and become aware of their infection, receive a diagnosis, and move to compartment $d$. The time to diagnosis is different for self-tests and non-self-tests;
2. may die.

Once diagnosed and in group $d$, a person may leave this group through death.

**Figure 1:** *A flow chart showing the progression described with the compartmental model*

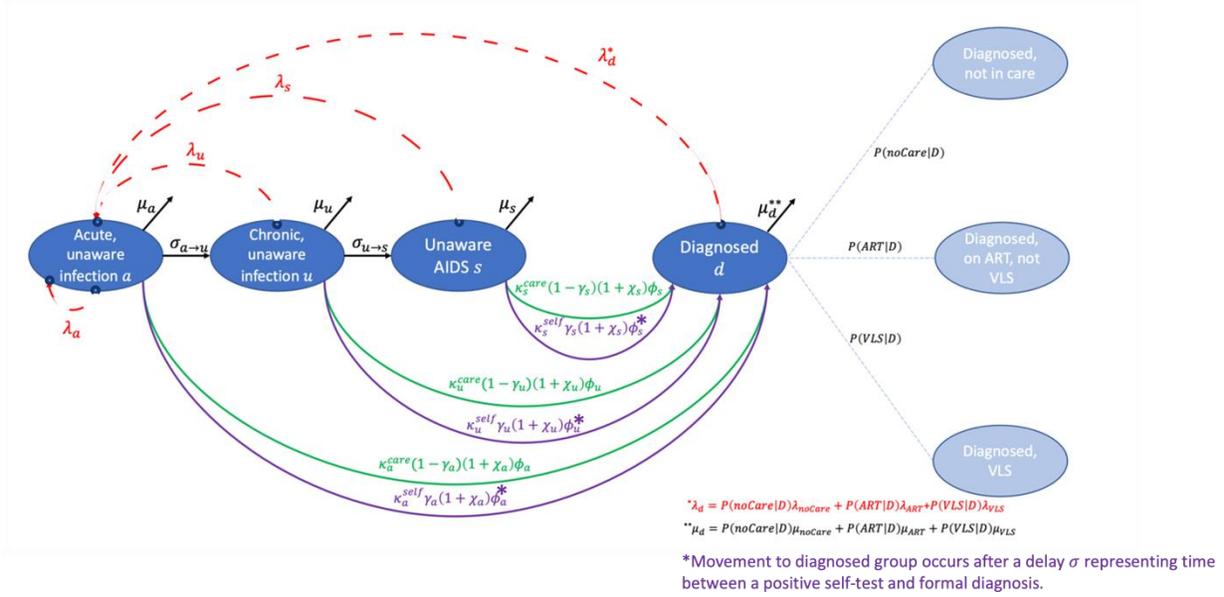

*Movement to diagnosed group occurs after a delay $\sigma$ representing time between a positive self-test and formal diagnosis.

Mathematically, this process is described with the following set of linear differential equations:

$$\begin{aligned}
\dot{a} &= \lambda_a a + \lambda_u u + \lambda_s s + \lambda_d d - (\sigma_{a\to u} + \widetilde{\phi_a} + \mu_a)a \\
\dot{u} &= \sigma_{a\to u} a - (\sigma_{u\to s} + \widetilde{\phi_u} + \mu_u)u \quad (1)\\
\dot{s} &= \sigma_{u\to s} u - (\widetilde{\phi_s} + \mu_s)s \\
\dot{d} &= \widetilde{\phi_a} a + \widetilde{\phi_u} u + \widetilde{\phi_s} s - \mu_d d,
\end{aligned}$$

where:

- $\lambda_i$ are the stage ($i$) specific transmission rates
- $\sigma_{i\to j}$ are the natural disease progression rates from $i \to j$ (note $\sigma_{a\to u}$ also corresponds to the testing *detection period*),
- $\mu_i$ are the stage ($i$) specific mortality rates, and
- $\widetilde{\phi_i}$ are the stage ($i$) specific detection rates, which describe the rate at which unaware PWH become aware of their status (see eq. (2) in the following section for mathematical definition).

Note that we do not model the non-PWH (susceptible) population explicitly. The justification of this choice is based on the low prevalence of HIV in the general population in the United States. We provide additional mathematical arguments for the validity and accuracy of this modeling choice, as well as several other important model properties, in Supplement A.

Before proceeding to a detailed discussion of model parameters, however, we remark that not modeling the non-PWH population explicitly affects the interpretation of those parameters, as they specifically refer to rates *among the PWH population.* This is particularly important for testing-related parameters, as the provided testing rates are based on testing rates among persons with undiagnosed HIV infection, and the disease-stage specific subpopulations therein. In general, these rates will differ from those of the general, non-PWH population. Hence, when changes in testing rates are discussed in the analyses that follow, these refer to changes in the testing rate among persons with undiagnosed HIV. While these may be related to larger changes in testing levels among non-PWH, the exact nature of this relationship is not specified in this model. This may potentially affect interpretation of the model results from a programmatic perspective; further elaboration is provided in the discussion section. Finally, it naturally follows that similar care must be taken when interpreting parameters defining mortality and transmission.

With these considerations in mind, the description of each parameter and the associated units are provided in Tables 1a-1c. We now discuss some important considerations regarding the testing, transmission, and mortality parameters. A more comprehensive discussion of model parameterization is provided in Supplement B.

*Modeling and parameterization of HIV testing*

We now briefly discuss how testing is modeled and parameterized.

The *detection rate* $\widetilde{\phi_i}$ for a compartment $i$ (recalling that all compartments refer to PWH at different stages) is defined as:

$$\widetilde{\phi_i} = \kappa_i^{self} \phi_i^{self} + \kappa_i^{care} \phi_i^{care}, \quad (2)$$

where $\kappa_i^{self}$ and $\kappa_i^{care}$ are the *test sensitivities* (unitless) for self and non-self-tests. We do not refer to this rate as a testing rate, as the sensitivities are, in general, less than one. This rate thus does not reflect all tests performed among PWH, but only those which yield a positive test, as reflected by the incorporation of test sensitivity. For a compartment $i$, and $\phi_i^{self}, \phi_i^{care}$, denote the corresponding *testing rates,* which may be either positive or negative tests, and accordingly, do not necessarily result in a diagnosis. Note that we consider non-self-tests to be gold standard laboratory tests, as reflected by the parameter values in Table 1. We acknowledge that this is a simplification, as, in general, non-self-tests may include rapid CLIA-waived tests conducted in community-based settings and other types of testing. Such tests are, however, a minority of non-self-tests [15]. We discuss this assumption further at the end of the current subsection.

Let $\phi_i$ represent the *testing* rate of the PWH population in the absence of self-testing hereafter referred to as the *baseline testing rate* for compartment $i$. We note the testing rate $\phi_i$ related to, but distinct from, the detection rate $\widetilde{\phi_i}$ defined in equation (2). The difference between the two rates is that the detection rate $\widetilde{\phi_i}$ accounts for test sensitivity, whereas the testing rate $\phi_i$ does not. Thus, while the testing rate $\phi_i$ defines the rate of tests in the PWH population in group $i$, the detection rate defines the rate of *positive tests*. Let $\chi_i$ then denote the percent change in the overall PWH testing rate in compartment $i$ due to self-testing, and $\gamma_i$ the proportion of tests among compartment $i$ that are self-tests. Then we can write the detection rate (2) as:

$$\widetilde{\phi_i} = \kappa_i^{self} \gamma_i (1 + \chi_i) \phi_i + \kappa_i^{care} (1 - \gamma_i)(1 + \chi_i) \phi_i. \quad (3)$$

In addition to sensitivity, self-tests and other tests may also differ in the time to receive a formal diagnosis after an initial positive result [13] . This difference is attributed to the need for further confirmatory laboratory testing after a positive self-test, delaying diagnosis and initiation-of-care. We modify the detection rate (2) above to account for this delay among PWH in compartment $i$, denoted as $t_i^{self \to diag.}$:

$$\widetilde{\phi_i} = \kappa_i^{self} \gamma_i \left[ \frac{1}{\left( \frac{1}{(1+\chi_i)\phi_i} + t_i^{self \to diag.} \right)} \right] + \kappa_i^{care}(1-\gamma_i)(1+\chi_i)\phi_i \quad (4).$$

Since we are considering non-self-tests as laboratory tests, the parameter is zero in this case, as we can assume diagnosis occurs at the same time as a positive result.

We may then rewrite the system in terms of (4) as:

$$\dot{a} = \lambda_a a + \lambda_u u + \lambda_s s + \lambda_d d - \left( \sigma_{a \to u} + \kappa_a^{self} \gamma_a \left( ((1+\chi_a)\phi_a)^{-1} + t_a^{self \to diag.} \right)^{-1} + \kappa_a^{care}(1-\gamma_a)(1+\kappa_a)\phi_a + \mu_a \right) a$$

$$\dot{u} = \sigma_{a \to u} a - \left( \sigma_{u \to s} + \left( ((1+\chi_u)\phi_u)^{-1} + t_u^{self \to diag.} \right)^{-1} + \kappa_u^{care}(1-\gamma_u)(1+\chi_u)\phi_u + \mu_u \right) u$$

$$\dot{s} = \sigma_{u \to s} u - \left( \left( ((1+\chi_s)\phi_s)^{-1} + t_s^{self \to diag.} \right)^{-1} + \kappa_s^{care}(1-\gamma_s)(1+\chi_s)\phi_s + \mu_s \right) s$$

$$\dot{d} = \kappa_a^{self} \gamma_a \left( ((1+\chi_a)\phi_a)^{-1} + t_a^{self \to diag.} \right)^{-1} a + \kappa_a^{care}(1-\gamma_a)(1+\chi_a)\phi_a a$$

$$+ \kappa_u^{self} \gamma_u \left( ((1+\chi_u)\phi_u)^{-1} + t_u^{self \to diag.} \right)^{-1} u + \kappa_u^{care}(1-\gamma_u)(1+\chi_u)\phi_u u$$

$$+ \kappa_s^{self} \gamma_s \left( ((1+\chi_s)\phi_s)^{-1} + t_s^{self \to diag.} \right)^{-1} s + \kappa_s^{care}(1-\gamma_s)(1+\chi_s)\phi_s s - \mu_d d.$$

The complete list of parameters, their names and units are provided in Table 1a-c. Full details on how these parameters are defined, based on surveillance data, are provided in Supplement B.

We acknowledge that assuming all non-self-tests are laboratory tests does not reflect the true testing landscape, which includes many point-of-care and community-based rapid testing programs [15]. This assumption is motivated by simplicity, as there are large jurisdictional differences in the availability of such programs and a relative paucity of data regarding their reach [15].

*Computation of mortality and transmission rates*

As mentioned previously, the compartment model does not directly simulate PWH through the continuum-of-care post-diagnosis. This modeling choice is motivated by the specific problem, as we assume that testing does not modify the rates of transitioning through the continuum-of-care after a diagnosis has been made. However, diagnosed PWH may still transmit HIV, and the rates at which such transmission occurs depend heavily on post-diagnosis continuum-of-care. Indeed, the uptake and adherence of antiretroviral therapy (ART) medications to achieve viral suppression (VLS) has been found to be the most significant factor in reducing HIV incidence [16], with PWH who are VLS carrying essentially no risk of sexual transmission [17], [18]. Hence, the post-diagnosis continuum of care is relevant for incidence estimation and was implicitly included in its estimation as follows.

We assume that the post diagnosis continuum-of-care distribution does not differ among PWH identified with self-test versus non-self-tests. We then consider three post-diagnosis subgroups:

1. PWH who are diagnosed and not on ART $d_{noCare}$,
2. PWH on ART but who are not virally suppressed $d_{ART}$,
3. PWH who are ART and virally suppressed $d_{VLS}$.

We can then consider the overall transmission rate in the post-diagnosis group as a weighted-average of the above groups. Mathematically, letting $\lambda_{noCare}, \lambda_{ART}, \lambda_{VLS}$ be the transmission rates corresponding to the listed groups, we can express this weighted average as:

$$\begin{aligned}\lambda_d d &= \lambda_{noCare} d_{noCare} + \lambda_{ART} d_{ART} + \lambda_{VLS} d_{VLS} \\ &= \lambda_{noCare} P(noCare|D)d + \lambda_{ART} P(ART|D)d + \lambda_{VLS} P(VLS|D)d \\ &= \left(\lambda_{noCare} P(noCare|D) + \lambda_{ART} P(ART|D) + \lambda_{VLS} P(VLS|D)\right) d,\end{aligned}$$

where the $P(stage|D)$ represent the conditional probability of being in the specified sub-stage, given that one has received a diagnosis, and $d$ is from (1), representing the diagnosed state. The weighted-average transmission rate is then obtained as:

$$\lambda_d = \lambda_{noCare} P(noCare|D) + \lambda_{ART} P(ART|D) + \lambda_{VLS} P(VLS|D).$$

This allows us to account for the effects of the post-diagnosis continuum-of-care without requiring its explicit simulation. We stress that this inherently assumes no difference in post-diagnosis care resulting from self-testing.

The mortality rate among diagnosed PWH $\mu_d$ can also be similarly defined as a weighted average:

$$\mu_d = \mu_{noCare} P(noCare|D) + \mu_{ART} P(ART|D) + \mu_{VLS} P(VLS|D).$$

Modeling the post-diagnosis continuum of care with weighted averages allows us to maintain model parsimony, simplify model analysis and implementation, and reduce the level of necessary input data without neglecting the relevant dynamics. In Fig. 2, we expand the flow chart from Fig. 1, detailing the implicit treatment of the post-diagnosis continuum-of-care. Values used for transmission-related parameters are given in Table 1b, and mortality-related parameters in 1c. Full details discussion on how these parameters are defined from data are given in Supplement B.

The introduced model offers several advantages over existing high-fidelity models. Accounting for the HIV continuum-of-care and susceptible population implicitly through weighted averages greatly reduces the necessary level of input data, allows for straightforward parameterization of the model directly from surveillance data, and results in minimal computational overhead (thousands of distinct model configurations simulated in seconds on conventional hardware). Additionally, the parsimony of our model makes it amenable to formal mathematical analysis (see Supplement A) and the derivation of clear threshold conditions for which we may expect self-testing to reduce HIV incidence and/or increase awareness of infection status among PWH.

**Figure 2:** *Visualization of how the post-diagnosis continuum-of-care is considered implicitly through weighted-averaging.*

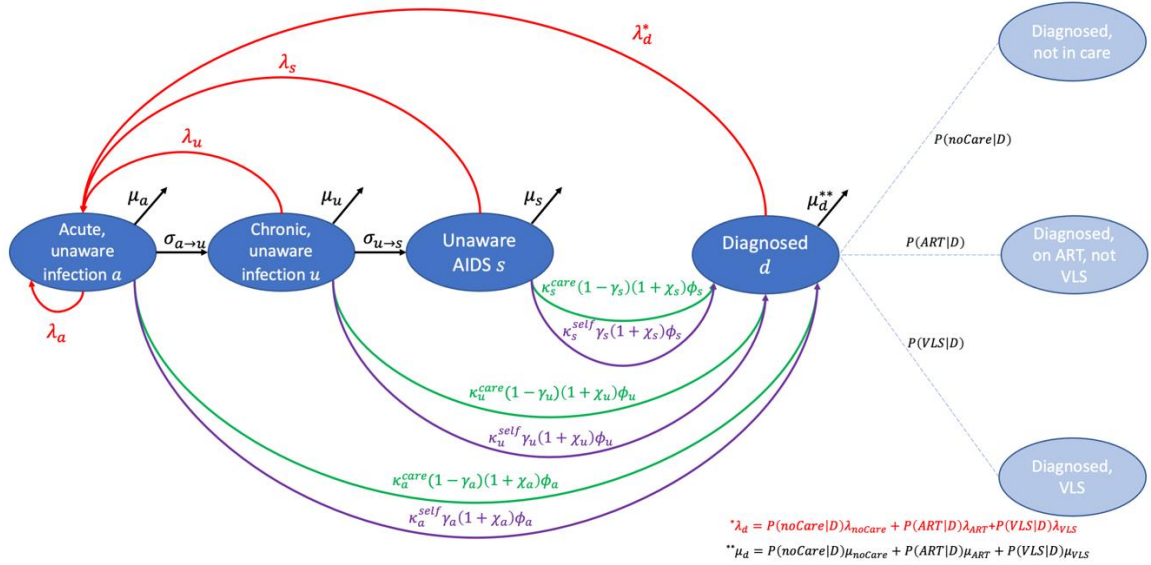

*Model validation: Comparison against national-level PATH3.0 model*

To validate the introduced model, we compared its performance against the PATH 3.0 model [19], a comprehensive stochastic agent-based model of HIV in the United States, calibrated to, and validated against, several epidemiological and care parameters in the National HIV Surveillance System (NHSS) [20], [21]. Briefly, we found that the compartment model matched PATH3.0 both quantitatively and qualitatively over a variety of metrics. Most importantly, the two models behave similarly in response to changes in levels of testing and self-testing. Full details are provided in Supplement C.

**Table 1a:** Input, fixed parameter values

| Parameter | Name | Value | Source |
|---|---|---|---|
| $\sigma_{a \to u}$ | Rate of movement from acute to chronic infection | 1/60 days | [22] |
| $\sigma_{u \to s}$ | Rate of movement from chronic infection to AIDS | 1/11.8 years | [23] |
| $t_s^{self \to diag.}$ | Delay in diagnosis after positive self-test, AIDS | 1/30 days | Assumed |
| $\phi_{a,u}$ | Baseline testing rate for PWH with acute and chronic infection | .0118 / months | Calculated using Surveillance data, 2017-19 [24]; see supplement B |
| $\phi_s$ | Testing rate for PWH with AIDS (late-stage) | .0481/months | Calculated using Surveillance data, 2017-19 [24]; set as $v_s \phi_u$ see supplement B |
| $v_s$ | Multiplier for increase in testing from AIDS compared to acute/chronic infection | 4.08 | [25] |
| $\kappa_a^{care}$ | Sensitivity of laboratory test to acute infection | .83 | [22] |
| $\kappa_a^{self}$ | Sensitivity of self-test to acute infection | 0.0 | [6] |
| $\kappa_{a,u}^{care}$ | Sensitivity of laboratory test to chronic infection | 1.0 | [26] |
| $\kappa_{u,s}^{self}$ | Sensitivity of self-test to chronic infection | .92 | [6] |

**Table 1b:** Incidence-related parameter values

| Parameter | Name | Value | Source |
|---|---|---|---|
| $\lambda_u$ | Transmission rate for chronic infection | .0074/months | Surveillance data, 2017-19 [24]; see supplement B |
| $\lambda_a$ | Transmission rate for acute infection | .0427/months | Surveillance data, 2017-19 [24], $\alpha_a$, see supplement B |
| $\alpha_a$ | Factor change in transmission probability for acute compared to chronic infection | 5.8 | [19] |
| $\lambda_s$ | Transmission rate for AIDS PWH | .0074/months | Surveillance data, 2017-19, [24]$\alpha_s$; see supplement B |
| $\alpha_s$ | Factor change in transmission probability for AIDS compared to chronic infection | 1.0 | Assumed |
| $\lambda_d$ | Transmission rate for diagnosed infection | .0019/months | Surveillance data, 2017-19 [24], $\alpha_{noCare}, \alpha_{ART}, \alpha_{VLS}$, see supplement B |
| $\lambda_{noCare}, \lambda_{ART}$ | Transmission rate for diagnosed PWH not care, diagnosed PWH on ART but not VLS | .0057/months | Surveillance data, 2017-19 [24], $\alpha_{noCare}, \alpha_{ART}$, see supplement B |
| $\alpha_{noCare}, \alpha_{ART}$ | Factor change in transmission probability for diagnosed, not in care and on ART< not VLS compared chronic infection | 0.78 | [19]; assumed equal. |
| $\lambda_{VLS}$ | Transmission rate for diagnosed PWH who are VLS | 0.0 | Surveillance data, 2017-19 [24], $\alpha_u$, see supplement B |
| $\alpha_{VLS}$ | Factor change in transmission for PWH who are VLS compared to chronic infection | 0.0 | [19] |

**Table 1c:** Mortality-related parameter values

| Parameter | Name | Value | Source |
|---|---|---|---|
| $\mu_a$ | Mortality rate, PWH with acute infection | .0069/years | Surveillance data, 2017-19 [24]; see supplement B |
| $\mu_u, \mu_{noCare}, \mu_{ART}$ | Mortality rate, PWH with chronic infection, diagnosed not in care, on ART but not VLS | .0174/years | Surveillance data, 2017-19 [24]; set as $\beta_u\mu_a$, $\beta_{noCare}\mu_a$, $\beta_{ART}\mu_a$; see supplement B |
| $\beta_u, \beta_{noCare}, \beta_{ART}$ | Factor change in mortality for PWH with chronic infection, diagnosed but not in care, and ART but not VLS compared to acute | 2.538 | [19], assumed equal. |
| $\mu_s$ | Mortality rate, PWH with AIDS, not diagnosed | .046/years | [19], see supplement B, set as $\beta_s\mu_a$ |
| $\beta_s$ | Factor change in mortality for PWH with undiagnosed AIDS compared to PWH with acute infection | 6.172 | [19] |
| $\mu_d$ | Mortality rate, PWH with diagnosed infection | .0086/years | Surveillance data, 2017-19 [24]; see supplement B |
| $\mu_{VLS}$ | Mortality rate, PWH who are VLS | .0043/years | Surveillance data, 2017-19 [24]; set as $\beta_{VLS}\mu_a$, see supplement B |
| $\beta_{VLS}$ | Factor change in mortality for PWH who are VLS compared to acute infection | 0.6346 | [19] |

**Analyses**

In this section, we apply the model to evaluate three different scenarios for HIV self-testing among MSM in the United States:

1. *Perfect supplementation:* If self-testing is added to existing testing levels and there is no replacement of non-self-tests with self-tests among PWH, we examine how much self-testing can reduce HIV incidence and increase awareness of status among the MSM population in the United States.
2. *Replacement analysis:* This assumes that when self-testing is expanded, some amount of replacement of non-self-tests with self-tests occurs among MSM PWH. We seek to determine, for varying levels of replacement, how much overall testing rates must increase among undiagnosed MSM PWH (MSM/PWH) to offset potentially negative effects of replacement, so that the net result supports incidence reduction.
3. *Sensitivity to self-test sensitivity and detection period:* Finally, we examine how changes to either self-test sensitivity or the self-testing detection period may affect results from the previous scenarios.

In each analysis, we parameterize the model using values (or methods) reported in Table 1a-1c. For the mortality and force-of-infection terms, we use a data-driven procedure to parameterize the model directly from public, national-level surveillance data for MSM in the United States provided on NCHHSTP ATLAS Plus [24]. These data are provided and summarized in Table 2, and the parameterization procedure is outlined in Supplement B.

After parameterization, the model undergoes a seven-year dry run before the validation period (years 2017-19). Figure 3 compares simulation values to the corresponding values computed from surveillance data, spanning the years 2017-19. Transmission rate is computed as $\frac{New\ HIV\ infections}{Total\ PWH\ population}$, while mortality rate is computed as $\frac{PWH\ Deaths}{Total\ PWH\ population}$. Self-testing is introduced at the beginning of the year 2020, and outcomes are collected over the years 2020-30.

We are interested in how self-testing affects both HIV incidence and awareness of status among MSM. We report the *percent change in incidence* as the relative difference in cumulative incidence $I$ from 2020-30 as compared to a *baseline scenario* with no self-testing ($\chi = 0$). Mathematically, this is defined as:

$$\%Incidence\ Change = \frac{\int_{2020}^{2030} \left(I_{self-test}(t) - I_{baseline}(t)\right) dt}{\int_{2020}^{2030} I_{baseline}(t) dt}.$$

For awareness of status, we report the percentage of MSM PWH who are diagnosed at the end of the modeled period (year-end 2030).

Before further discussion of our results, we briefly note that we restricted our analysis to MSM populations for two key reasons. First, MSM comprised 67% of new HIV diagnoses and 65.7% of estimated new HIV transmissions in the U.S. in 2021; thus, is a priority population for expanding HIV prevention efforts and reducing HIV incidence in the U.S. [27], [28]. Second, given the simple, four-compartment, single-population structure of the introduced model, restricting the analysis to this important subgroup helps simplify important questions regarding differences in awareness levels, testing rates and transmission behaviors across different populations.

**Table 2:** *Surveillance data used to parameterize model, United States MSM (source: NCHHSTP AtlasPlus [24]). Transmission rate calculated as (est. new HIV infections)/(est. PWH population). Mortality rate calculated as (PWH mortality)/(Est. PWH population)*

| Year | 2017 | 2018 | 2019 | Mean |
|---|---|---|---|---|
| **HIV diagnoses** | 25,345 | 24,464 | 23,870 | 24,560 |
| **HIV prevalence (including unaware)** | 655,100 | 673,000 | 689,900 | 672,667 |
| **HIV incidence** | 24,500 | 23,900 | 23,100 | 23,833 |
| **HIV deaths** | 7,010 | 6,888 | 7,184 | 7,027 |
| **% aware of status** | 83.8% | 84.3% | 84.8% | 84.3% |
| **% linked to care (among diag.)** | 77.0% | 77.3% | 77.7% | 77.3% |
| **% VLS (among diag.)** | 66.1% | 67.3% | 68.1% | 67.1% |
| **New cases / new deaths** | 3.50 | 3.47 | 3.22 | 3.39 |
| **Transmission rate [yr$^{-1}$]** | 0.037 | 0.035 | 0.033 | 0.036 |
| **Mortality rate [yr$^{-1}$]** | 0.011 | 0.010 | 0.010 | 0.010 |
| **New diagnoses/Total undiagnosed [yr$^{-1}$]** | 0.231 | 0.226 | 0.220 | 0.226 |

**Figure 3:** *Comparison of relevant rates, surveillance (2017-19) compared to simulation (2017-19). Data taken from NCHHSTP AtlasPlus. Transmission rate calculated as (est. new HIV infections)/(est. PWH population). Mortality rate calculated as (PWH mortality)/(Est. PWH population).*

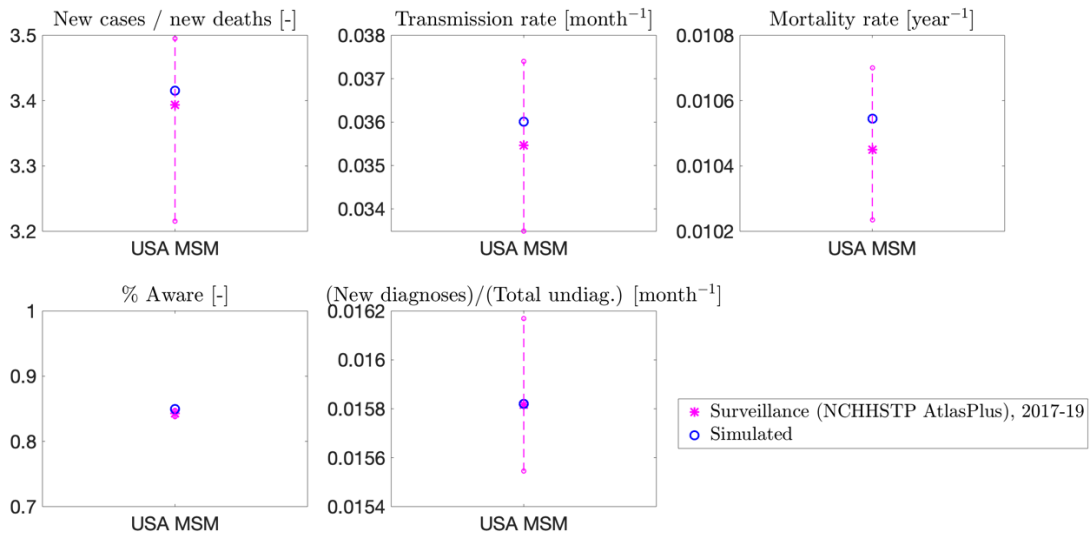

**Analysis 1: Perfect supplementation**

To demonstrate the potential favorable effects of HIV self-testing on HIV incidence and awareness of status among MSM, we first consider the special case of perfect supplementation. In this scenario, self-tests are assumed to not be a replacement of non-self-tests among undiagnosed PWH, and 100% of the increase in the overall testing rate among undiagnosed PWH is attributable to adding self-testing to the existing non-self-testing efforts. Mathematically, this corresponds to the condition:

$$\gamma = \frac{\chi}{1+\chi}.$$

At the programmatic level, this can be interpreted as an optimistic situation, in which self-testing increases the testing rate among MSM who do not test regularly, while MSM who already test regularly do not change their testing behavior.

We vary the time period of formal diagnosis and engagement following a positive self-test as $t^{self \rightarrow diag.}$= 1, 2, 3, 6, and 12 months. As the amount of non-self-testing does *not* change in these scenarios, we expect that no incidence increases will occur. However, the amount that incidence decreases will depend on both the extent of supplementation and how quickly formal diagnosis and engagement in care occurs, post-self-test.

**Figure 4:** Incidence reduction (left) and awareness of status (right) among US MSM for varying levels of self-test supplementation and time to formal diagnosis $t^{self \rightarrow diag.}$

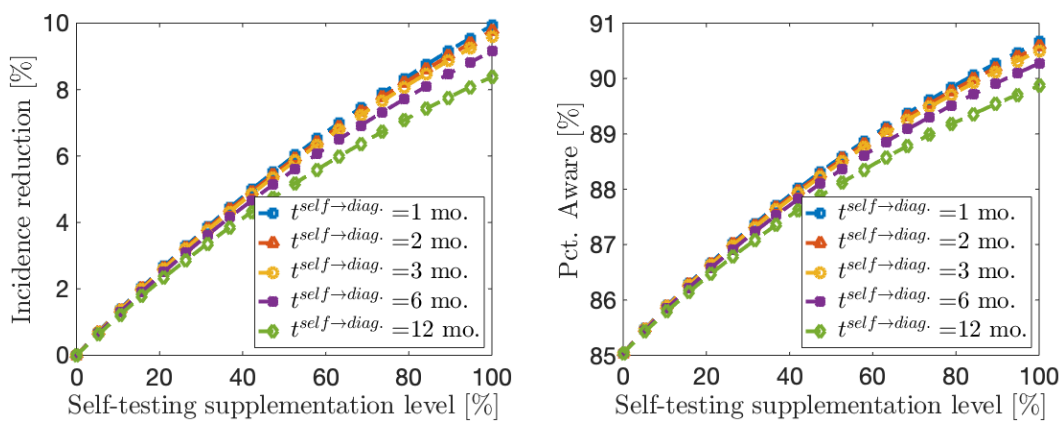

*Results*

We plot the results of the analysis in Figure 4. As expected, no incidence increases occur. For levels of supplementation under ~40%, we saw similar incidence decreases and levels of awareness for different levels of time-to-formal diagnosis $t^{self \rightarrow diag.}$. However, as the amount of supplementation increased, the importance of obtaining a formal diagnosis and promptly initiating care became more apparent. In the case of 100% supplementation, $t^{self \rightarrow diag.} = 1$ month, resulted in a 10% decrease in incidence and nearly 91% of MSM PWH aware of their status, compared to an 8% incidence decrease and just under 90% of MSM PWH aware of status when time to formal diagnosis $t^{self \rightarrow diag.} = 12$ months. Considering the incidence estimates shown in Table 2, these results correspond to about 2400 and 1800 fewer annual new infections, respectively, compared to the baseline scenario.

This scenario showed that, assuming no replacement occurs, increased HIV self-testing among MSM PWH may provide an effective way to reduce HIV incidence. To maximize potential incidence reductions, it is important to ensure that formal diagnosis and prompt engagement in care occur after a positive self-test. Nonetheless, this effect is small within this scenario of perfect supplementation, and only changed incidence by 2% in the most extreme case.

**Analysis 2: Replacement analysis**

The perfect supplementation scenario shown above illustrated the best possible case for expanding HIV self-testing, in which testing is increased among MSM who do not test regularly as a result of the added self-testing, while regularly-testing MSM do not change their testing behavior. In reality, expanded self-testing may result in some replacement of non-self-tests with HIV self-tests among MSM PWH. Some persons who test regularly may switch to using self-tests with no change in the overall testing rate (*replacement*). However, simultaneously, some MSM who rarely or never test may begin to test more often, and further, some persons who test regularly increase their rate of testing, due to expanded self-testing (*supplementation*). This analysis sought to understand how much supplementation among MSM PWH must occur to offset the effects of replacement and ensure that self-testing does not result in incidence increases in that population – we call this the *threshold* level of supplementation.

We considered 25,000 random samples of in which the change in overall testing rate $\chi$ and percentage of tests that are self-tests $\gamma$ were varied independently from 0 to 100%. For each sample of 25,000 cases, the time to formal diagnosis after a positive self-test $t^{self \rightarrow diag.}$ was varied as 1, 2, 3, 6, and 12 months. We assumed a *baseline scenario* as one with no self-testing ($\gamma = 0$), and the overall calibrated MSM PWH testing rate over 2017-2019 ($\chi = 0$). Note that with no self-testing ($\gamma = 0$), time to formal diagnosis $t^{self \rightarrow diag.}$ has no effect (this parameter does not influence testing outside of self-tests), and hence the baseline scenario was identical for all $t^{self \rightarrow diag.}$. For each $t^{self \rightarrow diag.} = 1, 2, 3, 6, 12$ months, we sought to determine the threshold level of supplementation for each given level of self-testing prevalence.

*Results*

We provide parameter plane plots in Figures 5a-5d, depicting changes in incidence from baseline (left) and overall awareness of status (right) for different scenarios. The x-axes correspond to increases in overall testing levels from baseline ($\chi$) and the y-axes, to the percentage of overall testing that are self-tests ($\gamma$). From top-to-bottom, the plots depict time to formal diagnosis $t^{self \rightarrow diag.}$ =1, 3, 6, and 12 months. The solid curve corresponds to the *threshold curve:* points to the left of this curve in the ($\gamma, \chi$) plane result in incidence increases from baseline, while points to the right of the curve result in incidence decreases. We refer to the area of the region to the left of the threshold curve as the *negative outcomes region.*

The dashed curve in Figures 5a-5d notes the *perfect supplementation curve,* discussed in the previous section. Points on this curve correspond to situations in which self-tests supplement the baseline testing scenario perfectly, with no replacement of non-self-tests with self-tests. Points to the right of the perfect supplementation curve correspond to situations in which increases in overall testing among PWH are *greater* than the level of self-testing (i.e., increases in the rate of non-self-testing among PWH). In these scenarios, even if self-testing had zero sensitivity, we would expect incidence reduction to occur. For this reason, we consider the perfect supplementation scenario (studied in the previous section) to represent the "best case scenario" for self-testing.

For shorter time to formal diagnosis $t^{self \rightarrow diag.}$, the area to the left of the threshold curve was small, showing that small increases in testing levels were sufficient to offset replacement effects. As $t^{self \rightarrow diag.}$ increased, the area to the left of this curve increased, and larger increases in testing levels were needed to offset the effects of replacement. This is further shown in Table 3 and Figure 6, where we report the threshold testing increase level $\chi$ for different levels of self-testing prevalence $\gamma$ for each $t^{self \rightarrow diag.}$, as well as the area of the negative outcomes region. These results show the importance of rapid engagement in care after positive self-test.

Summary results showing the results in terms of incidence changes and overall awareness of status are provided in Table 4. The importance of quickly obtaining a formal diagnosis and engaging in care, post self-test, is again apparent; a shorter time to formal diagnosis $t^{self \rightarrow diag.}$ resulted in large decreases in incidence and higher levels of HIV status awareness. Note these aggregate outcomes did not consider cases to the right of the perfect supplementation curve, for reasons discussed previously.

This study demonstrated several important findings. First, ensuring smaller values for $t^{self \rightarrow diag.}$, that is, that formal diagnosis occurs quickly after a positive self-test, was important for both minimizing the risk of incidence increases, as well as maximizing potential incidence decreases. The results showed significant variation based on this factor. Second, provided that the time to formal diagnosis $t^{self \rightarrow diag.}$ is short (under ~3 months) and replacement levels among PWH are not excessive, the necessary increases in testing levels to offset replacement were not large; they were significantly smaller than the level of self-testing. In other words, for a given relative prevalence of self-testing $\gamma$, the corresponding threshold testing increase $\chi$ necessary to reduce incidence was much smaller than $\gamma$ ; for $\gamma = .25$, testing increases of less than 4% or less were sufficient, provided $t^{self \rightarrow diag.}$ was under 3 months.

**Table 3:** Threshold $\chi$ (increase in testing levels from baseline among MSM PWH) and negative outcomes region, for different levels of self-testing $\gamma$. This tells us, for a given percentage of self-testing among overall testing, how much increase in baseline testing rate was necessary to avoid replacement effects and ensure incidence decreases. The negative outcomes region refers to the area to the left of the threshold curve in the $(\chi, \gamma)$ plane, which result in incidence increases.

| $t^{self \rightarrow diag.}$ | Threshold $\chi$ $\gamma = .25$ | Threshold $\chi$ $\gamma = .5$ | Threshold $\chi$ $\gamma = .75$ | Threshold $\chi$ $\gamma = 1.0$ | Area, neg. outcome region |
|---|---|---|---|---|---|
| 1 month | 3.5% | 7.3% | 11.4% | 15.6% | 0.075 |
| 2 months | 3.8% | 7.9% | 12.5% | 17.2% | 0.082 |
| 3 months | 4.1% | 8.5% | 13.6% | 18.9% | 0.089 |
| 6 months | 4.9% | 10.4% | 16.8% | 23.9% | 0.110 |
| 12 months | 6.3% | 13.9% | 23.7% | 35.7% | 0.155 |

**Table 4:** Summary indicators for incidence changes, awareness of status, and negative outcomes among MSM/PWH in the replacement analysis.

| $t^{self \rightarrow diag.}$ | Max % incidence decrease (2020-30) | Max % incidence increase (2020-30) | Avg % incidence change (2020-30) | Min % Aware (2030) | Max % aware (2030) | Avg % aware (2030) |
|---|---|---|---|---|---|---|
| 1 month | -9.9% | 2.2% | -4.6% | 83.8% | 90.7% | 87.8% |
| 2 months | -9.7% | 2.4% | -4.5% | 83.7% | 90.6% | 87.7% |
| 3 months | -9.6% | 2.6% | -4.3% | 83.5% | 90.5% | 87.6% |
| 6 months | -9.1% | 3.0% | -3.8% | 83.2% | 90.2% | 87.3% |
| 12 months | -8.4% | 4.0% | -2.9% | 82.6% | 89.8% | 86.8% |

**Figure 5a-c:** Results of the replacement analysis for time to formal diagnosis $t^{self \to diag.} = 1$, 6, and 12 months (top-to-bottom). The solid line depicts the *threshold curve;* scenarios to the right of this curve will ensure reductions in incidence. The dashed curve depicts *perfect supplementation with no replacement*- this can be considered the "best case scenario" for self-testing; scenarios to the right of this curve correspond to increases in testing greater than the level of self-testing (hence, increases in non self-testing).

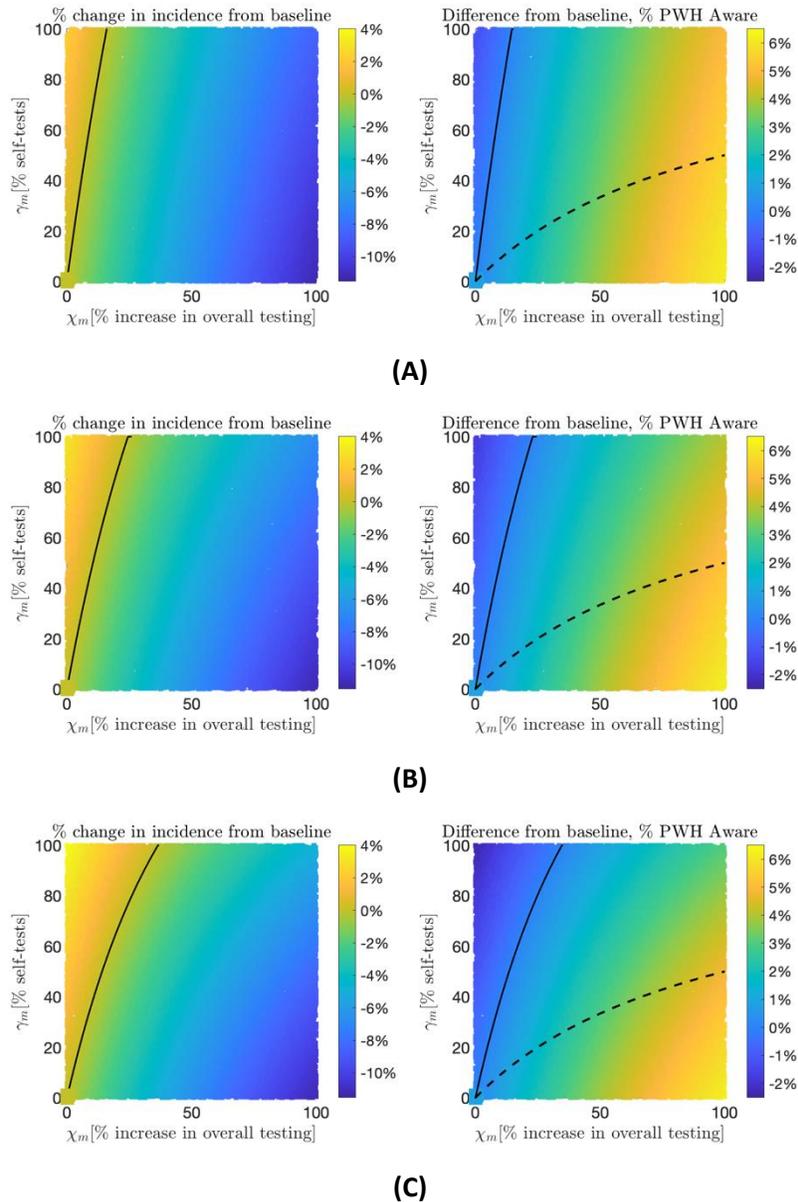

(A)

(B)

(C)

**Figure 6** Threshold $\chi$ (increase in testing levels from baseline among MSM PWH) for different percentages of self-testing $\gamma$ and time to formal diagnosis $t^{self \to diag.}$. The solid line denotes the threshold $\chi$; the dashed line denotes the perfect supplementation scenario.

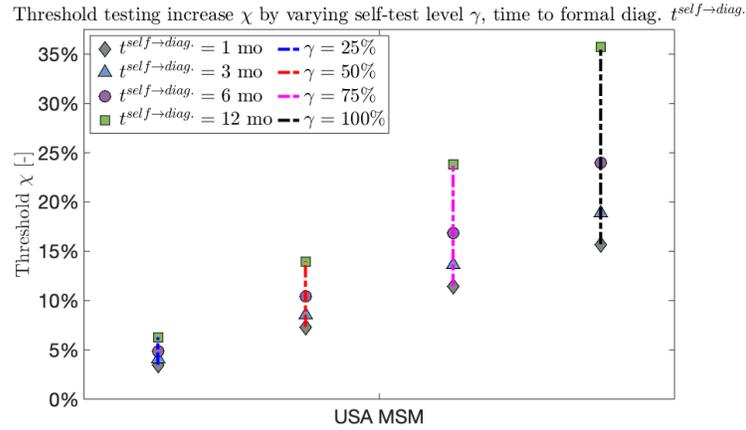

Threshold testing increase $\chi$ by varying self-test level $\gamma$, time to formal diag. $t^{self \to diag.}$

### Analysis 3: sensitivity to self-test sensitivity, detection period

We now evaluate the effect of self-test sensitivity $\kappa_{u,s}^{self}$, and detection period $\sigma_{a \to u}$. In the previous section, following the FDA guidelines, we considered self-test sensitivity $\kappa_{u,s}^{self} = .92$, and that the period necessary for self-tests to become accurate (detection period) was governed by $\sigma_{a \to u} = 60$ days. These values may be conservative, and some literature suggests that detection period $\sigma_{a \to u}$ may be as low as 42 days, or as high as 90 days, and that self-test sensitivity $\kappa_{u,s}^{self}$ may be as high as .95 [29], [30], [31], [32].

To analyze the effects of these parameters on our conclusions, we performed a sensitivity analysis. We considered the same underlying setup as in the preceding scenarios. We let self-test sensitivity $\kappa_{u,s}^{self} = .92, .94, .96$, and, for each $\kappa_{u,s}^{self}$, we performed simulations for detection period $\sigma_{a \to u} = 30, 60, 90$ days, in line with the range of estimates in the literature. As in previous scenarios, we performed simulations for $t^{self \to diag} = 1, 2, 3, 6$ and 12 months.

*Results*

We found that both self-test sensitivity $\kappa_{u,s}^{self}$ and detection period $\sigma_{a \to u}$ had small effects on threshold $\chi$ levels for lower levels of self-test prevalence $\gamma$. However, for higher levels of self-testing, the impact of these two values increases (Figure 7). The magnitude of effect for each parameter is similar.

We also examined the effects of the test sensitivity $\kappa_{u,s}^{self}$ and detection period $\sigma_{a \to u}$ on the area of the negative outcomes region and average percent incidence decrease. In Figure 8, we see that the effects of changes in each parameter on the negative outcomes region are quite consistent, and result in a near-constant change to the region area, despite differences in the other parameters. Decreasing the detection period $\sigma_{a \to u}$ from 90 to 30 days reduced the area of the negative outcome region by approximately .035, across the range of self-test sensitivity $\kappa_{u,s}^{self}$ and time to formal diagnosis $t^{self \to diag.}$ values. Similarly, increasing test sensitivity $\kappa_{u,s}^{self}$ from .92 to .96 led to a corresponding decrease of approximately .035 (Tables 5a-5b) across the range of detection period $\sigma_{a \to u}$ and time to formal diagnosis $t^{self \to diag.}$ values.

The analyses imply that the effects changes of in detection period or self-test sensitivity on the negative outcomes region are independent of each other, and insensitive on time to formal diagnosis $t^{self \to diag.}$.

We analyzed the effect of self-test sensitivity $\kappa_{u,s}^{self}$ and detection period $\sigma_{a \to u}$ on mean percent incidence decrease. As depicted in Figure 8, increasing self-test sensitivity from .92 to .96 showed a modest effect, decreasing incidence by approximately 0.4-0.5% (Tables 5c-5d). This decrease appeared insensitive to different values of both detection period $\sigma_{a \to u}$ and time to formal diagnosis $t^{self \to diag}$. We found the reducing the detection period $\sigma_{a \to u}$ had a smaller effect compared to test sensitivity $\kappa_{u,s}^{self}$, particularly at lower levels of time to formal diagnosis $t^{self \to diag}$, on mean percent incidence decrease (Tables 5c-5d). When formal diagnosis occurred after $1 - 2$ months, varying the detection period $\sigma_{a \to u}$ only reduced incidence by around 0.1%. When the time from a positive self-test to formal diagnosis was longer, we observed a larger, though still small effect, and decreasing the detection period $\sigma_{a \to u}$ from 90 to 30 days reduced incidence by about 0.3%.

Overall, we found that both reducing the detection period and increasing test sensitivity led to small reductions in the increases in testing necessary to offset replacement effects. Additionally, improvements in these areas may result in slightly larger incidence decreases from expanded self-testing. Test sensitivity showed a larger effect on incidence compared to the detection period. The observed effects showed little significant interdependence, or dependence on possible delays in formal diagnosis and care initiation after a positive self-test ($t^{self \to diag}$). Further, they were substantially smaller than the effects of $t^{self \to diag}$ on its own. As such, potential improvements in these areas did not change the qualitative nature of the previous analyses; improvements will result in small reductions in the necessary testing increases to ensure incidence reduction, and further reductions in incidence.

**Figure 7:** We plot the necessary testing increase (threshold $\chi$) for different levels of self-testing prevalence ($\gamma$) for different self-test sensitivities ($\kappa_{u,s}^{self}$), detection/window periods ($\sigma_{a \to u}$). Note that $t_{self \to diag} = 3$ months in each case.

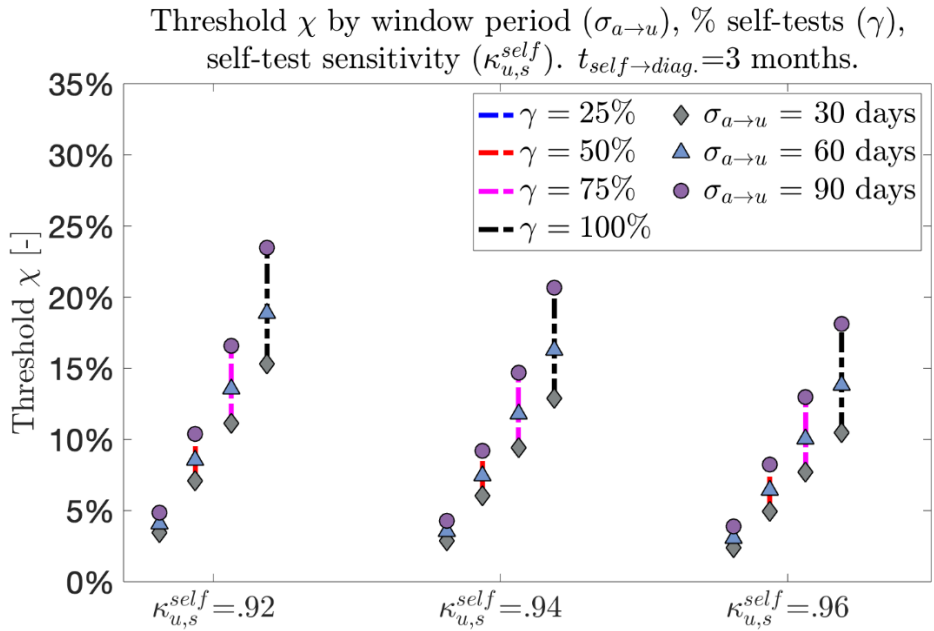

**Figure 8:** We plot the area of the negative outcomes region (left) and mean % incidence decrease over 2017-30 (right) as a function of $t^{self \to diag}$ for different detection/detection periods $\sigma_{a \to u}$ and self-test sensitivities $\kappa_{u,s}^{self}$.

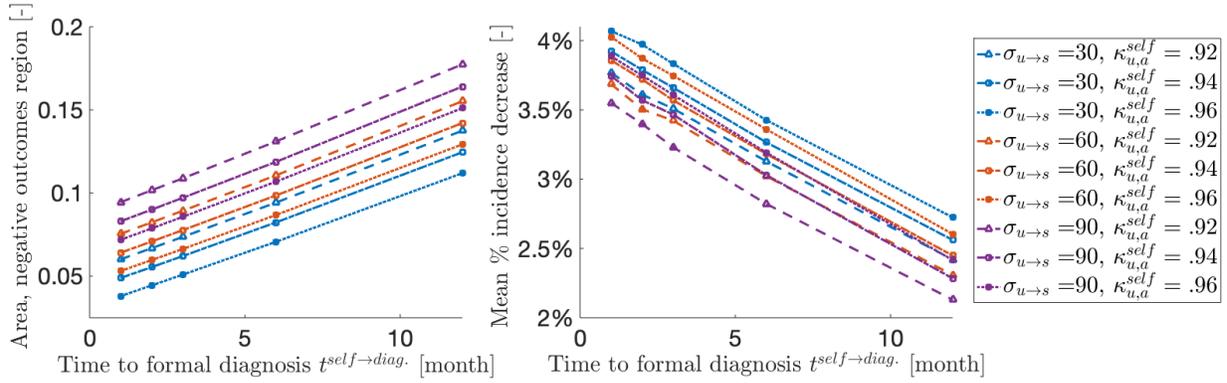

**Table 5a:** Area of the negative outcomes region (2020-30) for different self-test sensitivities ($\kappa_{u,s}^{self}$) and detection periods ($\sigma_{a \to u}$). Time to formal diagnosis ($t^{self \to diag}$)=3 months.

|  | $\sigma_{a \to u}$ = 30 days | $\sigma_{a \to u}$ = 60 days | $\sigma_{a \to u}$ = 90 days |
|---|---|---|---|
| $\kappa_{u,s}^{self}$ = .92 | .073 | .089 | .108 |
| $\kappa_{u,s}^{self}$ = .94 | .062 | .077 | .097 |
| $\kappa_{u,s}^{self}$ = .96 | .051 | .066 | .085 |

**Table 5b:** Area of the negative outcomes region (2020-30) for different self-test sensitivities ($\kappa_{u,s}^{self}$) and detection periods ($\sigma_{a \to u}$). Time to formal diagnosis ($t^{self \to diag}$)=12 months.

|  | $\sigma_{a \to u}$ = 30 days | $\sigma_{a \to u}$ = 60 days | $\sigma_{a \to u}$ = 90 days |
|---|---|---|---|
| $\kappa_{u,s}^{self}$ = .92 | .137 | .154 | .176 |
| $\kappa_{u,s}^{self}$ = .94 | .124 | .141 | .163 |
| $\kappa_{u,s}^{self}$ = .96 | .111 | .128 | .150 |

**Table 5c:** Simulated mean % incidence decrease (2020-30) of the for different self-test sensitivities ($\kappa_{u,s}^{self}$) and detection periods ($\sigma_{a \to u}$). Time to formal diagnosis ($t^{self \to diag}$)=3 months.

|  | $\sigma_{a \to u}$ = 30 days | $\sigma_{a \to u}$ = 60 days | $\sigma_{a \to u}$ = 90 days |
|---|---|---|---|
| $\kappa_{u,s}^{self}$ = .92 | -4.4% | -4.3% | -4.1% |
| $\kappa_{u,s}^{self}$ = .94 | -4.6% | -4.5% | -4.3% |
| $\kappa_{u,s}^{self}$ = .96 | -4.8% | -4.7% | -4.5% |

**Table 5d:** Simulated mean % incidence decrease, 2020-30 for different self-test sensitivities ($\kappa_{u,s}^{self}$) and detection periods ($\sigma_{a \to u}$). Time to formal diagnosis ($t^{self \to diag}$)=12 months.

|  | $\sigma_{a \to u}$ = 30 days | $\sigma_{a \to u}$ = 60 days | $\sigma_{a \to u}$ = 90 days |
|---|---|---|---|
| $\kappa_{u,s}^{self}$ = .92 | -3.0% | -2.9% | -2.7% |
| $\kappa_{u,s}^{self}$ = .94 | -3.2% | -3.1% | -2.8% |
| $\kappa_{u,s}^{self}$ = .96 | -3.4% | -3.3% | -3.1% |

**Discussion**

We introduced a parsimonious 4-compartment model to quantify the effect of self-testing on HIV transmission and awareness of HIV status. By focusing on the pre-diagnosis period in detail and considering the post-diagnosis continuum of care only implicitly, we reduced model complexity, while still describing our problem of interest.

We then applied the model to examine the case of perfect supplementation, in which self-testing strictly adds to existing testing among MSM with undiagnosed HIV, with no replacement. We found that in the supplementation scenario, self-testing can decrease incidence by as much as 10%. While formal diagnosis and engagement-in-care occurring promptly after diagnosis ensures the largest possible incidence reductions in the perfect supplementation scenario, this effect is small (a 2% difference in incidence).

We then performed an analysis in a more realistic scenario in which we allowed replacement of non-self-tests with self-tests among MSM with undiagnosed HIV. This study showed that self-testing can provide consistent decreases in HIV incidence, even if some replacement occurs, provided formal diagnosis and engagement in care after a positive self-test occur quickly. Longer delays may necessitate larger increases in the testing rate to ensure incidence reduction. Further, ensuring such delays are short also leads to larger decreases in incidence.

Finally, we examined how possible differences in self-test performance, in terms of self-test sensitivity and self-test detection period, may change our conclusions. Improvements in either or both areas would further decrease incidence and reduce the possible negative impact of replacement. However, these effects were small.

Our study was not directly comparable to those performed in [13], [14] as we considered a national-level model , rather than individual cities. Nonetheless, our analysis was more optimistic about self-testing than [13], [14]. The analysis in [14] showed increases in incidence of nearly 50% for strict substitution of non-self-tests with self-tests, and potentially as much as 300% for long delays in formal diagnosis after a positive self test $t^{self \to diag}$. In [13], the case of 25% self-tests with no increase in overall testing rate increased incidence 0.28% in Atlanta and 3.75% in Seattle, while 50% self-tests and no increases in overall testing increased incidence 3.4% in Atlanta and 8.75% in Seattle. In contrast, our analysis never showed increases of more than 4.0% in incidence, even for the most extreme case of 100% self-testing, no increase in overall testing rate, and a 12-month delay in formal diagnosis and engagement in care. We also found that the potential reductions in incidence were also greater (as much as 10%) in our analysis compared to the 1% and 5% reductions found in [13]. Further, we found that the increases in overall testing levels necessary to offset any replacement effects among PWH with undiagnosed HIV are small, especially for rapid formal diagnosis and engagement in care after a positive self-test. We note that our study specifically considers changes in testing among MSM with HIV, rather than among the general MSM population. As such the modeled changes in testing behavior may not be directly comparable, and this may explain some or all the differences in our findings.

However, our analysis was consistent with the findings in [13], [14] regarding two important points: strict substitution of non-self-tests with self-tests had a negative effect on HIV incidence, and that ensuring that time to formal diagnosis and engagement in care after a positive self-test was the most important factor in guaranteeing that self-testing reduced incidence.

**Limitations**

We acknowledge several shortcomings of our analysis and modeling approach. The introduced compartmental model does not further stratify the population of interest into sub-groups within each compartment. This allows us to keep the necessary number of input parameters low, and to easily obtain parameter values directly from surveillance data.. However, testing patterns vary among individuals [4], [7], [8], [9], [10] and CDC recommends different testing intervals for groups based on background prevalence and individual characteristics [33], [34]. To help alleviate some of these concerns, we restricted the analysis to the MSM population in the current analysis. However, accounting for further behavioral differences within the MSM population, as was done in [13], [14], may improve model performance, provided reliable data can be obtained. This is particularly important for modeling interventions that prioritize groups known to have lower testing levels. We believe self-testing benefits would be even more likely among PWH populations or jurisdictions with lower background testing rates, as overall testing rates are more likely to increase when baseline testing rates are lower. Our analysis shown here, which exhausted the overall parameter space, does generally account for such scenarios, such as the perfect-supplementation analysis. However, it is difficult to define and study more precisely defined interventions without expanding the modeling framework. Extending the model resolution to account for such scenarios is unlikely to change our quantitative conclusions. However, such analyses may be useful for intervention planning.

Our model did not consider behavioral changes resulting from negative self-tests. As a result, this analysis does not address any potential concern arising from, for example, a false-negative self-test leading to increased transmission risk behavior. Additional data from HIV self-testing implementation studies, including data on harms, would help to refine this model.

Linearizing the transmission in our model also has its advantages, most notably the fast simulation (order of seconds) of thousands of scenarios and the way in which we may parameterize from surveillance data (see Supplement B). These characteristics enabled us to easily characterize outcomes over the entire parameter space and precisely identify threshold conditions. However, this linearization is made possible by assuming that HIV prevalence is relatively stable, allowing us to assume that changes in the susceptible fraction of the population are small enough to be ignored. Our formal analysis (Supplement A) shows that this assumption is valid, and that any error incurred by the linearization can be bounded in terms of model parameters. Given the epidemiologically-relevant ranges of these parameters, our bounds suggest that these assumptions are acceptable when considering the entire US MSM population. While this modeling approach is likely generalizable to some degree to entire countries/jurisdictions/populations with a similar HIV profile, it may be unsuited for analyses in countries/jurisdictions/populations with higher levels of HIV prevalence and/or transmission, as well as with rapidly changing demographic profiles.

These modeling assumptions also affect the practical interpretation of our results. Since non-PWH are not explicitly considered, all modeled changes in testing behavior refer to changes in testing among PWH with undiagnosed HIV. While most of the current HIV modeling literature either only explicitly model the undiagnosed PWH population as was done here or define relationships between the testing rates of undiagnosed PWH and non-PWH, this model does not define, or seek to define, any such relationship. This model could consider testing among the general population by assuming some functional relationship between testing among undiagnosed PWH and non-PWH without significant structural modification in future extensions of this model.    [25], [35], [36], [37]

The model also may be limited in its direct programmatic application. From the programmatic point of view, two distinct HIV testing strategies may lead to the same changes in testing within a particular population or community. For example, one testing strategy may increase existing testing levels 10% among that population homogenously, across the board, implying a 10% increase among those undiagnosed PWH within that population as well. However, a carefully designed HIV testing strategy may specifically aim to prioritize testing among those persons, communities, or settings where undiagnosed HIV may be more prevalent, that could result in a 10% testing increase among the undiagnosed PWH, but a much lower testing increase among non-PWH. The scope, design, and cost of the two HIV testing strategies would be different; however, the effect on overall awareness of status among MSM PWH, and future incidence, would be the same. The modeling approach used here, in the absence of any additional information, cannot distinguish between the two HIV testing strategies.

From a practical point of view, the analysis should be extended to Ending the HIV Epidemic (EHE) priority jurisdictions [38], [39]. We considered a national-level model of MSM as a proof-of-concept to demonstrate the overall potential benefit of HIV self-testing on HIV incidence among MSM in the U.S. However, different jurisdictions vary greatly in underlying levels of incidence, prevalence, HIV testing among PWH, and awareness of status. Accounting for populations besides MSM may also require that the model introduce additional stratifications. These changes may result in varying conclusions; self-testing may be expected to provide a larger or smaller effect on incidence, or different levels of testing increases may be necessary in different jurisdictions or among different populations. Any such potential differences would have important implications for future intervention efforts. This is an important direction for future work. We note that the modeling framework introduced here is well-suited for such a task, as it allows for easy, efficient parameterization from a small amount of easily obtained surveillance data.  An important next step is to apply this model to particular local jurisdiction scenarios where the local dynamics are taken into consideration and more direct recommendations could be made regarding how best to design an HIV self-testing program.

While we considered HIV self-testing in this work, we note that the modeling framework introduced here may be applied generically to compare any two (or more) different types of tests. To this end, such an approach may be applied to evaluate the efficacy of other types of tests  compared to the gold standard laboratory-based tests, such as rapid CLIA-waived tests in community settings.

Finally, in the present work, we considered a simplified testing situation, in which all non-self-tests were assumed to be laboratory tests. More realistic models should account for the heterogeneous testing landscape outside of self-testing and incorporate community-based and point-of-care rapid testing programs. From the modeling perspective, such effects would be reflected in differences in the sensitivity and time-to-diagnosis parameters for non-self-tests. Relative levels of laboratory, community-based, and point-of-care testing vary by jurisdiction [15], and incorporating such effects will likely require a jurisdiction-specific approach. In general, we may expect such effects to lead to reduced sensitivity for non-self-tests and increased delays in diagnosis after a positive result. Provided adequate input data, the modeling framework introduced here can account for these effects. Future work applying this model, especially at the jurisdiction-specific level, should consider these factors. However, we note that accounting for these heterogeneities would, in general, reduce self-testing threshold levels. Thus, the results shown here are conservative; actual incidence decreases may be even larger, and any necessary testing increases even smaller, than those shown here. Thus, the current results likely understate the positive impacts of expanding self-testing, particularly regarding the threshold levels.

**Conclusion**

Self-testing offers clear advantages over other forms of testing in terms of convenience and availability. Our modeling study of MSM in the United States assessed the concern that, given their lower sensitivity, self-tests could perhaps lead to overall increases in HIV incidence if self-tests replace laboratory-based testing. We found these concerns to be minimal, although some small risk remains in certain scenarios. For these reasons, it is important to expand HIV self-testing programs in a purposeful way that minimizes the possible risks while maximizing the potential benefits. From our analysis, we found that the most important factor in developing an effective self-testing program is ensuring that confirmatory laboratory-based testing and engagement in care occur promptly after a positive self-test. Additionally, to offset any potential replacement effects, programs must ensure that the expansion of self-testing results in increases in the underlying population testing rate. Focusing self-testing programs in jurisdictions or among populations with lower background testing rates would likely reach more people who have never or rarely ever tested, which could improve the chances of increasing overall testing rates, and thus reducing incidence. Finally, continued research on developing more sensitive HIV self-tests, particularly for recently acquired infection, should be pursued. Provided such conditions can be assured, expanded HIV self-testing can become an even more effective tool in the fight against HIV.

**Acknowledgments:** *The authors would like to acknowledge Kevin Delaney, Elizabeth DiNenno, John Brooks, Matthew Biggerstaff, Rebecca Borchering, Sinead Morris, Emily Pollock, Dina Mistry, Bruce Beau, and Angela Hutchinson for their helpful input and suggestions.*

**Supplement A**

**Mathematical analysis**

In this section, we consider the following nonlinear model, whose relation to the system (1) in the main text will be established in the following:

$$\dot{e} = \Lambda - \mu_e e - \left(\frac{\widetilde{\lambda_a}a}{n} + \frac{\widetilde{\lambda_u}u}{n} + \frac{\widetilde{\lambda_s}s}{n} + \frac{\widetilde{\lambda_d}d}{n}\right)e$$

$$\dot{a} = \left(\frac{\widetilde{\lambda_a}a}{n} + \frac{\widetilde{\lambda_u}u}{n} + \frac{\widetilde{\lambda_s}s}{n} + \frac{\widetilde{\lambda_d}d}{n}\right)e - (\sigma_{a \to u} + \widetilde{\phi_a} + \mu_a)a$$

$$\dot{u} = \sigma_{a \to u} a - (\sigma_{u \to s} + \widetilde{\phi_u} + \mu_u)u \qquad (A1)$$

$$\dot{s} = \sigma_{u \to s} u - (\widetilde{\phi_s} + \mu_s)s$$

$$\dot{d} = \widetilde{\phi_a} a + \widetilde{\phi_u} u + \widetilde{\phi_s} s - \mu_d d.$$

The parameters in (A1) are similar to those in (1) in the main text, with the addition of the compartment $e$ (for *eligible,* denoted as such to avoid confusion with the $s$ compartment) denotes the eligible (or susceptible) population, $n = e + a + u + s + d$ is the total living population, $\Lambda$ (units Persons/Time) is the rate-of-entry into the eligible population, and $\mu_e$ (units 1/Time) is the mortality rate of the eligible (susceptible) population. $\widetilde{\lambda_i}$, $i = a, u, s, d$ have the same function and units as $\lambda_i$ in main text (1), however, their values are different in general.

For ease of notation, we will also denote the *eligible (susceptible) population fraction* as:

$$\Sigma = \frac{e}{n}$$

We now prove several results regarding the system (A1).

**Proposition 1:** The system (1) in the main text is the *linearized infection subsystem* of (A1) linearized about $\Sigma_0 = \Sigma(t_0)$.

*Proof*: Following the procedure outlined in [40], [41], [42], we first note that, in (A1), the infection can be transferred by all compartments except $e$. We then define:

$$x = \begin{pmatrix} a \\ u \\ s \\ d \end{pmatrix}$$

And write (1) in the form:

$$\dot{x} = f(x) - v(x).$$

Where $f$ defines new infections and $v$ defines all other movement in the system.

By inspection of (A1), we find:

$$f(x) = \begin{pmatrix} (\widetilde{\lambda_a}a + \widetilde{\lambda_u}u + \widetilde{\lambda_s}s + \widetilde{\lambda_d}d)\Sigma \\ 0 \\ 0 \\ 0 \end{pmatrix}, v(x) = \begin{pmatrix} (\sigma_{a \to u} + \widetilde{\phi_a} + \mu_a)a \\ -\sigma_{a \to u}a + (\sigma_{u \to s} + \widetilde{\phi_u} + \mu_u)u \\ -\sigma_{u \to s}u + (\widetilde{\phi_s} + \mu_s)s \\ -\widetilde{\phi_a}a - \widetilde{\phi_u}u - \widetilde{\phi_s}s + \mu_d d \end{pmatrix}.$$

We now compute the *Jacobian matrices F and V of f and v*, respectively, evaluated at $\Sigma = \Sigma_0$:

$$F_{i,j} = \frac{df(x_i)}{dx_j}\Big|_{\Sigma=\Sigma_0}, V_{i,j} = \frac{dv(x_i)}{dx_j}\Big|_{\Sigma=\Sigma_0}.$$

Standard computations give:

$$F = \begin{pmatrix} \widetilde{\lambda_a} \times \Sigma_0 & \widetilde{\lambda_u} \times \Sigma_0 & \widetilde{\lambda_s} \times \Sigma_0 & \widetilde{\lambda_d} \times \Sigma_0 \\ 0 & 0 & 0 & 0 \\ 0 & 0 & 0 & 0 \\ 0 & 0 & 0 & 0 \end{pmatrix}, \quad (A2)$$

$$V = \begin{pmatrix} \sigma_{a \to u} + \widetilde{\phi_a} + \mu_a & 0 & 0 & 0 \\ -\sigma_{a \to u} & \sigma_{u \to s} + \widetilde{\phi_u} + \mu_u & 0 & 0 \\ 0 & -\sigma_{u \to s} & \widetilde{\phi_s} + \mu_s & 0 \\ -\widetilde{\phi_a} & -\widetilde{\phi_u} & -\widetilde{\phi_s} & \mu_d \end{pmatrix}. (A3)$$

The linearized infection subsystem about $\Sigma_0$ is then given by:

$$\dot{x} = (F - V)x,$$

letting $\lambda_i = \widetilde{\lambda_i} \times \Sigma_0$ for $i = a, u, s, d$, gives the system:

$$\begin{aligned} \dot{a} &= \lambda_a a + \lambda_u u + \lambda_s s + \lambda_d d - (\sigma_{a \to u} + \widetilde{\phi_a} + \mu_a)a \\ \dot{u} &= \sigma_{a \to u}a - (\sigma_{u \to s} + \widetilde{\phi_u} + \mu_u)u \\ \dot{s} &= \sigma_{u \to s}u - (\widetilde{\phi_s} + \mu_s)s \\ \dot{d} &= \widetilde{\phi_a}a + \widetilde{\phi_u}u + \widetilde{\phi_s}s - \mu_d d, \end{aligned} \quad (A4)$$

which is the same as the system (1) in the main text, completing the proof.

**Proposition 2:** The effective reproduction number $R_t$ for $\Sigma = \Sigma_0$ of the system (A1) and (A4) is given by:

$$R_t = \frac{\lambda_a}{(\sigma_{a \to u} + \widetilde{\phi_a} + \mu_a)} + \frac{\lambda_u \sigma_{a \to u}}{(\sigma_{a \to u} + \widetilde{\phi_a} + \mu_a)(\sigma_{u \to s} + \widetilde{\phi_u} + \mu_u)} + \frac{\lambda_s \sigma_{a \to u}\sigma_{u \to s}}{(\sigma_{a \to u} + \widetilde{\phi_a} + \mu_a)(\sigma_{u \to s} + \widetilde{\phi_u} + \mu_u)(\widetilde{\phi_s} + \mu_s)}$$
$$+ \frac{\lambda_d}{\mu_d}\left[\frac{\widetilde{\phi_a}(\sigma_{u \to s} + \widetilde{\phi_u} + \mu_u)(\widetilde{\phi_s} + \mu_s) + \widetilde{\phi_u}\sigma_{a \to u}(\widetilde{\phi_s} + \mu_s) + \widetilde{\phi_s}\sigma_{a \to u}\sigma_{u \to s}}{(\sigma_{a \to u} + \widetilde{\phi_a} + \mu_a)(\sigma_{u \to s} + \widetilde{\phi_u} + \mu_u)(\widetilde{\phi_s} + \mu_s)}\right]$$

*Proof:*

As detailed in [16], [40], [41], [42], $R_t$ is given by the largest eigenvalue of the *next generation matrix* $NV^{-1}$, where $N$ and $V$ are the matrices defining the linearized infection subsystem. In the proof of Proposition 1, we found that, for the system (A1) $N$ and $V$ are given by (A2) and (A3) respectively.

Computation yields:

$$V^{-1} = \begin{pmatrix} \frac{1}{V_{1,1}} & 0 & 0 & 0 \\ \frac{\sigma_{a \to u}}{V_{1,1}V_{2,2}} & \frac{1}{V_{2,2}} & 0 & 0 \\ \frac{\sigma_{a \to u}\sigma_{u \to s}}{V_{1,1}V_{2,2}V_{3,3}} & \frac{\sigma_{u \to s}}{V_{2,2}V_{3,3}} & \frac{1}{V_{3,3}} & 0 \\ \frac{\widetilde{\phi_a}}{V_{1,1}\mu_d} + \frac{\widetilde{\phi_u}\sigma_{a \to u}}{V_{1,1}V_{2,2}\mu_d} + \frac{\widetilde{\phi_s}\sigma_{a \to u}\sigma_{u \to s}}{V_{1,1}V_{2,2}V_{3,3}\mu_d} & \frac{\widetilde{\phi_u}}{V_{2,2}\mu_d} + \frac{\sigma_{u \to s}\widetilde{\phi_s}}{V_{2,2}V_{3,3}\mu_d} & \frac{\widetilde{\phi_s}}{V_{3,3}\mu_d} & \frac{1}{\mu_d} \end{pmatrix}$$

, where $V_{i,j}$ indicates the $i,j$-th entry of $V$.

Denoting the $i$-th column of $V^{-1}$ as $v_i$, and letting $n = (\lambda_a, \lambda_u, \lambda_s, \lambda_d)^T$, we immediately verify that:

$$NV^{-1} = \begin{pmatrix} n^T v_1 & n^T v_2 & n^T v_3 & n^T v_4 \\ 0 & 0 & 0 & 0 \\ 0 & 0 & 0 & 0 \\ 0 & 0 & 0 & 0 \end{pmatrix},$$

$R_t$ is then obtained as the *spectral radius* of $NV^{-1}$, denoted as $\rho(NV^{-1})$, the magnitude of the largest eigenvalue of $NV^{-1}$ [16], [40], [41], [42].

Observe that $NV^{-1}$ has only one non-zero row, implying $NV^{-1}$ has rank one, hence only one nonzero eigenvalue [43]. One immediately finds that the only eigenvector corresponding to a nonzero eigenvalue is $(1,0,0,0)^T$ with eigenvalue $n^T v_1$. It follows that:

$$|n^T v_1| = \rho(NV^{-1}) = R_t.$$

**Theorem 1:** At each $t$, the derivative of the susceptible population fraction:

$$\Sigma(t) = \frac{susceptible\ population\ (t)}{susceptible\ population\ (t) + infected\ population\ (t)} = \frac{e}{n}$$

in (A1) obeys the bound:

$$|\dot{\Sigma}| \leq \left(\frac{\Lambda}{n} + \widetilde{\lambda_a} + \mu_s - \mu_e\right)|1 - \Sigma|, \quad (A5)$$

Where the dependence of $\Sigma$, $\dot{\Sigma}$, and $n$ on time is understood if not denoted explicitly. Further, assuming $n \geq n_{min}$ for some $n$, for all $t$:

$$|\dot{\Sigma}| \leq \left(\frac{\Lambda}{n_{min}} + \widetilde{\lambda_a} + \mu_s - \mu_e\right). \quad (A6)$$

*Proof:* We consider the system (A1) and observe:

$$\frac{d}{dt}[\Sigma] = \frac{d}{dt}\left[\frac{e}{n}\right]$$
$$= \frac{\dot{e}n - \dot{n}e}{n^2}.$$

$\dot{e}$ is given by the first equation in (A1), and $\dot{n}$ by summing all the equations in (A1). This yields:

$$|\dot{\Sigma}| = \left| \frac{\left(\Lambda - \mu_e e - (\widetilde{\lambda_a} a + \widetilde{\lambda_u} u + \widetilde{\lambda_s} s + \widetilde{\lambda_d} d)\frac{e}{n}\right)n - (\Lambda - (\mu_e e + \mu_a a + \mu_u u + \mu_s s + \mu_d d))e}{n^2} \right|$$

$$= \left| \frac{\Lambda - \mu_e e - (\widetilde{\lambda_a} a + \widetilde{\lambda_u} u + \widetilde{\lambda_s} s + \widetilde{\lambda_d} d)\frac{e}{n}}{n} + \frac{(\mu_e e + \mu_a a + \mu_u u + \mu_s s + \mu_d d - \Lambda)e}{n^2} \right|$$

$$\leq \left| \frac{\Lambda - \mu_e e}{n} + \frac{(\mu_e e + \mu_a a + \mu_u u + \mu_s s + \mu_d d - \Lambda)e}{n^2} \right| + \left| \frac{(\widetilde{\lambda_a} a + \widetilde{\lambda_u} u + \widetilde{\lambda_s} s + \widetilde{\lambda_d} d)e}{n^2} \right|.$$

Noting that $\left|\frac{e}{n}\right| \leq 1$ by definition, $\widetilde{\lambda_a} \geq \widetilde{\lambda_{u,s,d}}$ in general, and $a + u + s + d = n - e$:

$$|\dot{\Sigma}| \leq \left| \frac{\Lambda - \mu_e e}{n} + \frac{(\mu_e e + \mu_a a + \mu_u u + \mu_s s + \mu_d d - \Lambda)e}{n^2} \right| + \left| \frac{\widetilde{\lambda_a}(n-e)}{n} \right|$$

$$= \left| \frac{\Lambda - \mu_e e}{n} + \frac{(\mu_e e + \mu_a a + \mu_u u + \mu_s s + \mu_d d - \Lambda)e}{n^2} \right| + \widetilde{\lambda_a} \left|1 - \frac{e}{n}\right|.$$

Again exploiting that $\left|\frac{e}{n}\right| \leq 1$, $a + u + s + d = n - e$, and observing that $\mu_s > \mu_{e,a,u,d}$ and $\mu_e < \mu_{a,u,d,s}$ in general:

$$|\dot{\Sigma}| \leq \left| \frac{\Lambda(n-e)}{n^2} + \frac{(\mu_e e + \mu_a a + \mu_u u + \mu_s s + \mu_d d - \mu_e n)e}{n^2} \right| + \widetilde{\lambda_a} \left|1 - \frac{e}{n}\right|$$

$$= \left| \frac{\Lambda(n-e)}{n^2} \right| + \left| \frac{\mu_e e + \mu_a a + \mu_u u + \mu_s s + \mu_d d - \mu_e(e + a + u + s + d)}{n} \right| + \widetilde{\lambda_a} \left|1 - \frac{e}{n}\right|$$

$$= \frac{\Lambda}{n}\left|1 - \frac{e}{n}\right| + \left| \frac{(\mu_a - \mu_e)a + (\mu_u - \mu_e)u + (\mu_s - \mu_e)s + (\mu_d - \mu_e)d}{n} \right| + \widetilde{\lambda_a} \left|1 - \frac{e}{n}\right|$$

$$\leq \frac{\Lambda}{n}\left|1 - \frac{e}{n}\right| + \left| \frac{(\mu_s - \mu_e)(n-e)}{n} \right| + \widetilde{\lambda_a} \left|1 - \frac{e}{n}\right|$$

$$= \left( \frac{\Lambda}{n} + \widetilde{\lambda_a} + \mu_s - \mu_e \right) \left|1 - \frac{e}{n}\right|,$$

which establishes (A5). The claim (A6) follows immediately by noting that $0 \leq \frac{e}{n} \leq 1$ and $\frac{\Lambda}{n} < \frac{\Lambda}{n_{min}}$ by definition.

*Corollary of Theorem 2:* The difference between the true susceptible fraction at a time $t_n$ and the linearized susceptible fraction at time $t_0$ is bounded by:

$$|\Sigma(t_n) - \Sigma_0| \leq \left( \frac{\Lambda}{n(t^*)} + \lambda_a + \mu_s - \mu_e \right) \left(1 - \Sigma(t_n^*)\right) (t_n - t_0) \qquad (A7)$$

with $t_0 \leq t_n^* < t_n$ for each $t_n$, and by:

$$\left( \frac{\Lambda}{n_{min}} + \lambda_a + \mu_s - \mu_e \right) (t_n - t_0) \qquad (A8)$$

for all $t_n$.

*Proof:* As both $e$ and $n$ are both at least once differentiable by definition and assumed nonzero, their quotient $\Sigma = \frac{e}{n}$ may be expanded in a Taylor series about $\Sigma_0$ for each $t_n$ such that:

$$\Sigma(t_n) = \Sigma_0 + \dot{\Sigma}(t_n^*)(t_n - t_0),$$

for some $t_n^*$ in $[t_0, t_n]$. From (A5):

$$|\Sigma(t_n) - \Sigma_0| = |\dot{\Sigma}(t_n^*)||(t_n - t_0)|$$
$$\leq \left(\frac{\Lambda}{n(t_n^*)} + \widetilde{\lambda_a} + \mu_s - \mu_e\right)|1 - \Sigma(t_n^*)|(t_n - t_0),$$

Establishing (A7). (A8) follows by an identical argument but applying (A6) rather than (A5).

Note that the bound (A8) is not a particularly sharp bound and can likely be improved, however, it is useful as it provides a uniform bound throughout the domain.

In practice, however, (A7) provides an effective bound. Generally, $\Sigma_0 \approx 1$, and, by the continuity of $\Sigma$, we then expect $|1 - \Sigma(t)| \approx 0$ for $t$ close to $t_0$, ensuring that the bound (A7) remains sharp for $t$ not excessively large.

We also note that (A7) and (A8) can likely be improved. $\widetilde{\lambda_a}$, while it provided a convenient upper bound for ease of analysis, is generally higher than the overall transmission rate, and not representative of its true value in practice. Additionally, the term $\frac{\Lambda}{n}$ is quite small in general, as $n$ is several orders of magnitude larger than $\Lambda$.

The bound (A7) is thus easy to interpret: it states that our linearized approximation of the system (A1) may remain quite accurate provided that:

- New entries to the eligible (susceptible) population are small compared to the overall population.
- Mortality and transmission remain low.
- The prevalence of HIV in the population is low.

The scenarios examined in the current document satisfy these conditions, and so we expect the linearized system (1) in the main text to provide a good approximation to (A1) throughout our time horizon.

**Supplement B**

*Compartment model parameterization*

In this section, we describe the procedure used to define the disease-stage specific transmission, mortality, and testing parameters introduced in the methods section of the main text. To inform our parameter definitions, we used measured data from ATLAS from the years 2017-2019 (for the simulations presented in the main text) and the PATH outputs from the baseline PATH simulation (for the PATH validation detailed in supplement C). In both instances, we used the full-population level transmission rate $\bar{\lambda}$, defined as:

$$\bar{\lambda} = \frac{Incidence}{Prevalence}.$$

averaged over the time-period. For the studies in the main text, this information, along with data regarding percentages of the population in care, aware, ART, and VLS, came from surveillance data provided by the ATLAS database [24]. We assume that, for any overall transmission level, the $\lambda_i$ are related by constant multiplicative factors $\alpha_i$. These factors were derived from [19] and reported in Table 1b in the main text.

We then calculated the individual $\lambda_i$ as:

$$\lambda_u = \frac{\bar{\lambda}}{\alpha_a P(A) + P(U) + \alpha_s P(S) + \alpha_{noCare} P(noCare) + \alpha_{ART} P(ART) + \alpha_{VLS} P(VLS)},$$
$$\lambda_a = \alpha_a \lambda_u,$$
$$\lambda_s = \alpha_s \lambda_u,$$
$$\lambda_{noCare} = \alpha_{noCare} \lambda_u,$$
$$\lambda_{ART} = \alpha_{ART} \lambda_u$$
$$\lambda_{VLS} = \alpha_{VLS} \lambda_{VLS}.$$

$P(ART), P(noCare), P(VLS)$ were available from our measured data and outputs (for the PATH validation) [19], [24].

$P(A), P(U), P(S)$, referring to the probabilities of being in the acute, chronic unaware, and AIDS unaware states, respectively, were obtained by first considering the probability that an individual is unaware of their HIV status $P(\bar{D})$, available from measurement and/or simulation:

$$P(\bar{D}) = 1 - P(D).$$

$P(A|\bar{D})$ and $P(S|\bar{D})$ were then determined through calibration. Accordingly:

$$P(A) = P(A|\bar{D}) \times P(\bar{D}), \quad P(S) = P(S|\bar{D}) \times P(\bar{D})$$

and

$$P(U) = P(\bar{D}) - P(A) - P(S).$$

Mortality rates were obtained analogously, with

$$\bar{\mu} = \frac{PWHDeaths}{Prevalence},$$

and the relevant data coming from either ATLAS (for the studies shown in the main text) or PATH model outputs (for the validation against PATH in supplement C). This approach yielded:

$$\mu_a = \frac{\bar{\mu}}{P(A) + \beta_u P(U) + \beta_s P(S) + \beta_{noCare} P(noCare) + \beta_{ART} P(ART) + \beta_{VLS} P(VLS)},$$
$$\mu_u = \beta_u \mu_a,$$
$$\mu_s = \beta_s \mu_a,$$
$$\mu_{noCare} = \beta_{noCare} \mu_a,$$
$$\mu_{ART} = \beta_{ART} \mu_a$$
$$\mu_{VLS} = \beta_{VLS} \mu_{VLS}.$$

The $\beta_i$ also came from [19] and are reported in Table 1c in the main text.

Testing rates (shown in Table 1a in the main text) were approximated with the following procedure.

First, using surveillance data, we define the following quantity:

$$\bar{\phi} \approx \frac{annual\ new\ diagnoses}{PWH\ unaware\ in\ previous\ year + annual\ new\ infection}.$$

Intuitively, this gives us the rate at which undiagnosed PWH receive a diagnosis. Note that, as it does not account for test sensitivity, it is distinct from the detection rate introduced in the methods section of the main text. However, it is the same quantity plotted in the bottom-center panel of Figure 3 in the main text.

Next, assume that undiagnosed PWH with chronic infection test at some unknown rate $\phi_u$, and further, that undiagnosed PWH with acute infection and late-stage infection (AIDS) test at $\nu_a$ and $\nu_s$ times the rate of undiagnosed PWH with chronic infection, respectively. Then $\phi_a = \nu_a \phi_u, \phi_s = \nu_s \phi_u$.

Let $\bar{D}$ refer to the probability of being an undiagnosed PWH. Then the probability of having undiagnosed acute and undiagnosed late-stage infection (AIDS) are respectively denoted as:

$$P(A|\bar{D}), \quad P(S|\bar{D}).$$

The probability of an undiagnosed PWH having a chronic-stage infection is then given by:

$$1 - P(A|\bar{D}) - P(S|\bar{D}).$$

Finally, we must also account for test sensitivity, assumed to vary by stage: $\kappa_{a,s,u}$.

The overall testing rate among undiagnosed PWH is then given by:

$$\phi_{undiag} = P(A|\bar{D})\phi_a + P(S|\bar{D})\phi_s + \left(1 - P(A|\bar{D}) - P(S|\bar{D})\right)\phi_u,$$

which, from our assumptions, reduces to:

$$\phi_{undiag} = \left(P(A|\bar{D})\nu_a + P(S|\bar{D})\nu_s + \left(1 - P(A|\bar{D}) - P(S|\bar{D})\right)\right)\phi_u.$$

Denote stage-specific test sensitivities as $\kappa_i$. If we assume that every positive test among an undiagnosed PWH leads to a diagnosis, then we have, approximately:

$$\overline{\phi} = \left(\kappa_a P(A|\overline{D})v_a + \kappa_s P(S|\overline{D})v_s + \kappa_u \left(1 - P(A|\overline{D}) - P(S|\overline{D})\right)\right)\phi_u,$$

implying:

$$\phi_u = \frac{\overline{\phi}}{\kappa_a v_a \alpha_a P(A|\overline{D}) + v_s \kappa_s \alpha_s P(S|\overline{D}) + \alpha_u \kappa_u \left(1 - P(A|\overline{D}) - P(S|\overline{D})\right)}.$$

We may then recover $\phi_{a,s}$ as:

$$\phi_a = v_a \phi_u, \quad \phi_s = v_s \phi_u.$$

For our studies in the main text, we assumed that $v_a = 1, v_s = 4.08$ (see [25]), $\alpha_{u,s} = 1, \alpha_a = .83$ to obtain our testing rates.

For the PATH validation study (shown in supplement C), since we tracked the number of tests performed in the PATH simulation, we defined $\phi_u$ as the base testing rates in PATH for undiagnosed PWH, with $\phi_s = v_s \phi_u, \phi_a = v_a \phi_u$.

**Supplement C**

**Model validation: Comparison against national-level PATH3.0 model**

To validate the introduced model, we compared its performance against the PATH 3.0 model [19], a comprehensive stochastic agent-based model of HIV in the United States, calibrated to, and validated against, several epidemiological and care parameters in the National HIV Surveillance System (NHSS) [20], [21]. The purpose of this study is to compare the compartment model to a validated, high-fidelity model of HIV transmission. In addition to verifying that the two models produce similar results for similar parameters, we aim to confirm that the models *respond* similarly to changes in self-testing related inputs (such as testing rate and self-testing prevalence). In this supplement, we provide the relevant details and results of this validation.

*Compartmental model setup*

The parameter values used to populate the compartment model are summarized in Table C1a-C1d, with appropriate source documentation and details provided. The parameter calibration process is explained in Supplement B. Briefly, all natural disease progression parameters are taken from the literature, and all care continuum related parameters are taken from the PATH simulation data. The remaining unknown parameters include the baseline testing rate, and transmission rate, which were calibrated using disease burden and care continuum data from PATH.

*PATH3.0 model setup*

To validate the introduced compartmental modeling, we compare the compartment model to PATH3.0, a stochastic, agent-based model. PATH3.0 has been previously validated by comparing multiple epidemic metrics against data from the National HIV Surveillance System for years 2010 to 2017. A full description of PATH3.0 and its relevant parameters can be found in [19].

The PATH model was modified to simulate self-testing behaviors. As in the compartment model, two parameters $\gamma$ and $\chi$ govern the percentage of self-tests and the change to the overall testing rate, respectively, with $t^{self \rightarrow diag.}$ giving the time to diagnosis after a positive self-test. The other new parameters are the self-test sensitivities for acute and chronic PWH and AIDS patients, $\kappa_a^{self}$ and $\kappa_{u,s}^{self}$. These are taken to be same as those in the compartment model and provided in Table C1a.

**Table C1a:** Input, fixed parameter values, validation against PATH3.0.

| Parameter | Name | Value | Source |
|---|---|---|---|
| $\sigma_{a \to u}$ | Rate of movement from acute to chronic infection | 1/60 days | [22] |
| $\sigma_{u \to s}$ | Rate of movement from chronic infection to AIDS | 1/11.8 years | [23] |
| $t_s^{self \to diag.}$ | Delay in diagnosis after positive self-test, AIDS | 1/30 days | Assumed |
| $\phi_{a,u}$ | Baseline testing rate for acute, chronic | .02/month | PATH rates |
| $\phi_s$ | Testing rate for AIDS stage. | .0816/month | PATH rates, set as $v_s \phi_{a,u}$ |
| $v_s$ | Multiplier for increase in testing from AIDS compared to acute/chronic infection | 4.08 | [25] |
| $\kappa_a^{care}$ | Sensitivity of laboratory test to acute infection | .83 | [22] |
| $\kappa_a^{self}$ | Sensitivity of self-test to acute infection | 0.0 | [6] |
| $\kappa_{a,u}^{care}$ | Sensitivity of laboratory test to chronic infection | 1.0 | [26] |
| $\kappa_{u,s}^{self}$ | Sensitivity of self-test to chronic infection | .92 | [6] |

**Table C1b:** Incidence-related parameter values, validation against PATH3.0

| Parameter | Name | Value | Source |
|---|---|---|---|
| $\lambda_u$ | Transmission rate for chronic infection | .0056/month | PATH outputs, calibration as detailed |
| $\lambda_a$ | Transmission rate for acute infection | .0325/month | PATH outputs, set as $\alpha_a \lambda_u$ |
| $\alpha_a$ | Factor change in transmission probability for acute compared to chronic infection | 5.8 | [19] |
| $\lambda_s$ | Transmission rate for AIDS PWH | .0056/month | PATH outputs, set as $\alpha_s \lambda_u$ |
| $\alpha_s$ | Factor change in transmission probability for AIDS compared to chronic infection | 1.0 | Assumed |
| $\lambda_d$ | Transmission rate for diagnosed infection | .0017/month | PATH outputs, calculation as detailed |
| $\lambda_{noCare}, \lambda_{ART}$ | Transmission rate for diagnosed PWH not care, diagnosed PWH on ART but not VLS | .0043/month | PATH outputs, set as $\alpha_{noCare}\lambda_u, \alpha_{ART}\lambda_u$ |
| $\alpha_{noCare}, \alpha_{ART}$ | Factor change in transmission probability for diagnosed, not in care and on ART< not VLS compared chronic infection | 0.78 | [19]; assumed equal. |
| $\lambda_{VLS}$ | Transmission rate for diagnosed PWH who are VLS | 0.000/month | PATH outputs, set as $\alpha_{VLS}\lambda_u$ |
| $\alpha_{VLS}$ | Factor change in transmission for PWH who are VLS compared to chronic infection | 0.0 | [19] |

**Table C1c:** Mortality-related parameter values, validation against PATH3.0

| Parameter | Name | Value | Source |
|---|---|---|---|
| $\mu_a$ | Mortality rate, acute infection | 0.0126 / years | PATH outputs, calibration as detailed |
| $\mu_u, \mu_{noCare}, \mu_{ART}$ | Mortality rate, chronic infection, diagnosed not in care, on ART but not VLS | 0.0320 / years | PATH outputs, set as $\beta_u\mu_a$, $\beta_{noCare}\mu_a$, $\beta_{ART}\mu_a$ |
| $\beta_u, \beta_{noCare}, \beta_{ART}$ | Factor change in mortality for PWH with chronic infection, diagnosed but not in care, and ART but not VLS | 2.538 | [19], assumed equal. |
| $\mu_s$ | Mortality rate, PWH with AIDS, not diagnosed | 0.0847 / years | [19], set as $\beta_s\mu_a$ |
| $\beta_s$ | Factor change in mortality for undiagnosed PWH with AIDS compared to acute infection | 6.172 | [19] |
| $\mu_d$ | Mortality rate, diagnosed infection | 0.0175 / years | PATH outputs, computed as detailed |
| $\mu_{VLS}$ | Mortality rate, PWH who are VLS | 0.0080 / years | PATH outputs, set as $\beta_{VLS}\mu_a$ |
| $\beta_{VLS}$ | Factor change in mortality for PWH who are VLS compared to acute infection | 0.6346 | [19] |

**Table C1d:** Continuum-of-care related parameter values, validation against PATH3.0

| Parameter | Name | Value | Source |
|---|---|---|---|
| $P(noCare|D)$ | Portion of diagnosed PWH not in care | .175 | [19], PATH outputs |
| $P(ART|D)$ | Portion of diagnosed PWH on ART, not VLS | .115 | [19], [24] PATH outputs |
| $P(VLS|D)$ | Portion of diagnosed PWH VLS | .710 | [19], [24] PATH outputs |
| $P(a|\overline{D})$ | Portion of undiagnosed who are acute | .0375 | Calibrated |
| $P(s|\overline{D})$ | Portion of undiagnosed with AIDS | .11 | Calibrated |
| $a_0$ | Initial acutely infected PWH | 14 | PATH initial conditions |
| $u_0$ | Initial chronic unaware PWH | 160 | PATH initial conditions |
| $s_0$ | Initial AIDS | 6 | PATH initial conditions |
| $d_0$ | Initial diagnosed PWH | 830 | PATH initial conditions |

*Scenarios*

We aim to see that the compartmental and PATH models respond similarly to changes in testing and self-testing levels. To this end, wee focused our analyses on simulating varying combinations of $\chi$ (% increase in testing rate from baseline), $\gamma$ (% of tests that are self-tests), and $t^{self \to diag.}$ (time from positive self-test to diagnosis), and estimating the corresponding changes in reach (proportion aware) and HIV incidence compared to a baseline scenario. We assume *baseline scenario as* $\gamma = 0$, $\chi = 0$; that is, no self-testing is present, and the overall testing rate is the calibrated population-level testing rate over 2017-2019. Note that with $\chi = 0$, $t^{self \to diag.}$ has no effect (this parameter does not influence testing outside of self-tests), and hence baseline case is identical for each $t^{self \to diag.}$. We evaluated $t^{self \to diag.} = 1, 2, 3, 6, 12$ months. For each $t^{self \to diag.}$, the $(\chi, \gamma)$- samples are drawn from $[0, 1] \times [0, 1]$; for the compartment model, we consider 25,000 random samples. For PATH3.0, we sample 121 pairs at the evenly-spaced points $0, 0.1, 0.2, \ldots, 1$ for both $\chi$ and $\gamma$.

For both models, we focused our analyses on men who have sex with men (MSM) given the higher prevalence in this group [44]. After a 12-year dry run beginning in 2006, we tracked incidence and the percentage of PWH aware of serological status among MSM from 2017-2030. For each $(\chi, \gamma)$ sample, the outputs of interest are the percent change from baseline in cumulative incidence from 2017-2030, and the overall % PWH who are aware of status at the end of the simulation period. For the compartment model, we also report the percent change in the effective reproduction number, $R_t$, from the baseline case. As the definition of $R_t$ in PATH3.0 is not straightforward, we do not provide this measure for the PATH3.0 simulations.

*Results*

We depict the baseline ($\chi = 0, \gamma = 0$) ratio of new infections to PWH deaths, Transmission rate, and mortality rates, by year, for the compartment and PATH models (Figure C3). Note the shown compartment model rates are not actually constant; however their smaller variation compared to PATH makes this difficult to observe. As PATH is a stochastic simulation, it is subject to more fluctuation than the deterministic compartment model. The mean values over the shown period for each indicator are given in Table C2.

**Fig. C3:** *new infection/mortality ratio, incidence and mortality rates calculated from simulation outputs for the PATH and compartment models.*

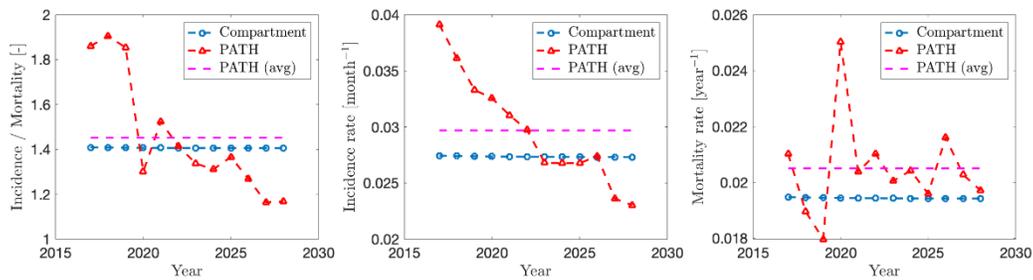

**Table C2:** *agreement in simulation outputs between PATH and compartment models for key baseline indicators over the simulated time period 2017-30.*

| Model | Avg. infection/death ratio | Avg. infection rate | Avg. mortality rate |
| --- | --- | --- | --- |
| PATH | 1.45 | .029/year | .0205/year |
| Compartment | 1.41 | .027/year | .0195/year |

In Fig. C4, we plot results from the validation study. The plots are arranged, from top to bottom, with $t^{self \to diag.} = 1, 2, 3, 6, 12$ months, respectively. From left-to-right, we plot percent change in incidence from baseline, the percent change in $R_t$ from baseline, and the percentage of PWH aware of their infection in the $(\chi, \gamma)$ plane, with each point corresponding to the respective sampled value.

The background of each plot shows the outputs of the compartmental simulations in the $(\chi, \gamma)$-plane, while the superimposed points above in the left and right plots show the outputs of the PATH simulation at the corresponding $(\chi, \gamma)$ point. The black curve in each plot delineates where the compartmental simulation changed sign, that is, the region in the $(\chi, \gamma)$-plane to the left of the black curve (hereafter referred to as the *negative outcomes region*) gave worsened outcomes compared to baseline. Conversely, to the right of the black curve, we saw reduced incidence and increased awareness of status for the respective $(\chi, \gamma)$. For the PATH simulation, points corresponding to improved outcomes (when compared to baseline) are plotted as circles, while the corresponding worsened outcomes are plotted as squares.

For both percent incidence change and awareness of status, we observe that the color-blending between the compartmental and PATH simulations is in good agreement, showing that the same $(\chi, \gamma)$ configurations yielded similar results for the two models. We also note that the square points (indicating points where PATH corresponds to worsened outcomes compared to baseline) are concentrated to the left of the black line, which marks the outcome delineation for the compartment model. This indicates that the models also were in broad agreement on the threshold points for which self-testing is beneficial.

Moving from top to bottom, we note that the threshold curve (and the corresponding square/circle split in the PATH simulations) was sensitive to changes in $t^{self \to diag.}$. For small $t^{self \to diag.}$, self-testing provided improved outcomes for all but the most extreme cases, in which self-testing simply replaced laboratory tests without changing the overall testing rate. However, as $t^{self \to diag.}$ grew larger, the threshold curve moved rightward, indicating a greater risk of negative outcomes. As the $t^{self \to diag.} = 6, 12$ month cases demonstrate, the overall testing rate had to increase to ensure that self-testing yielded a net positive effect. In Table C3, we provide the minimum, maximum and average percent changes in incidence, the baseline, minimum, maximum, and average levels of awareness of HIV status, and the area of the negative outcomes region for all simulations.

From this analysis, we draw two primary conclusions. The first is that the compartment model, despite its simplified, linearized structure, shows similar qualitative and quantitative behavior as compared to the PATH3.0 agent-based model. For similar baseline parameters, the models produce similar results. Furthermore, the two models respond similarly under changes to testing and self-testing levels. As PATH has been previously validated [19], from this agreement between the models, we can establish the validity of the compartmental model, giving us confidence in the results presented in the main text.

**Table C3:** Quantitative agreement in simulation outputs between PATH and Compartment model

| Model | $t^{self \to diag.}$ | Min % incidence change (2017-30) | Max % incidence change (2017-30) | Avg % incidence change (2017-30) | Baseline % aware (2030) | Min % Aware (2030) | Max % aware (2030) | Avg % aware (2030) | Area, neg. outcome region (2017-30) |
|---|---|---|---|---|---|---|---|---|---|
| Compartment | 1 month | -9.3% | 2.4% | -4.3% | 90.5% | 89.6% | 94.6% | 92.5% | 0.082 |
| PATH | 1 month | -8.8% | 2.8% | -4.2% | 90.2% | 88.2% | 96.1% | 93.1% | 0.082 |
| Compartment | 2 months | -9.3% | 2.7% | -4.2% | 90.5% | 89.4% | 94.6% | 92.4% | 0.091 |
| PATH | 2 months | -9.1% | 3.2% | -4.2% | 90.2% | 87.8% | 96.1% | 93.1% | 0.083 |
| Compartment | 3 months | -9.3% | 2.9% | -4.1% | 90.5% | 89.3% | 94.6% | 92.4% | 0.101 |
| PATH | 3 months | -8.6% | 2.1% | -4.1% | 90.2% | 88.4% | 96.1% | 93.0% | 0.124 |
| Compartment | 6 months | -9.3% | 3.7% | -3.7% | 90.5% | 89.0% | 94.6% | 92.2% | 0.130 |
| PATH | 6 months | -8.6% | 2.9% | -3.8% | 90.2% | 87.7% | 96.1% | 92.9% | 0.150 |
| Compartment | 12 months | -9.3% | 5.0% | -3.0% | 90.5% | 88.3% | 94.6% | 91.9% | 0.193 |
| PATH | 12 months | -8.6% | 3.4% | -3.1% | 90.2% | 87.7% | 96.1% | 92.8% | 0.190 |

*Fig. C4:* Simulation results of the PATH/Compartment validation. From top, $t_{a,u}^{self \to diag.} = 1, 2, 3, 6, 12$ months. The background are outputs from the compartmental simulations, while the colored points above are the PATH simulation outputs. From left to right, the plots show % incidence change from baseline over the simulated period 2017-30, % $R_t$ change from baseline over the simulated period 2017-30, and % of PWH aware of HIV status at the end of 2030 in the $(\chi, \gamma)$- plane. Black curves show where sign of compartment simulation changed (i.e., from increase to decrease). PATH simulations plotted as circles/squares corresponding to decrease/increase. Please refer to appropriate section in supplement text for a detailed interpretation of these results.

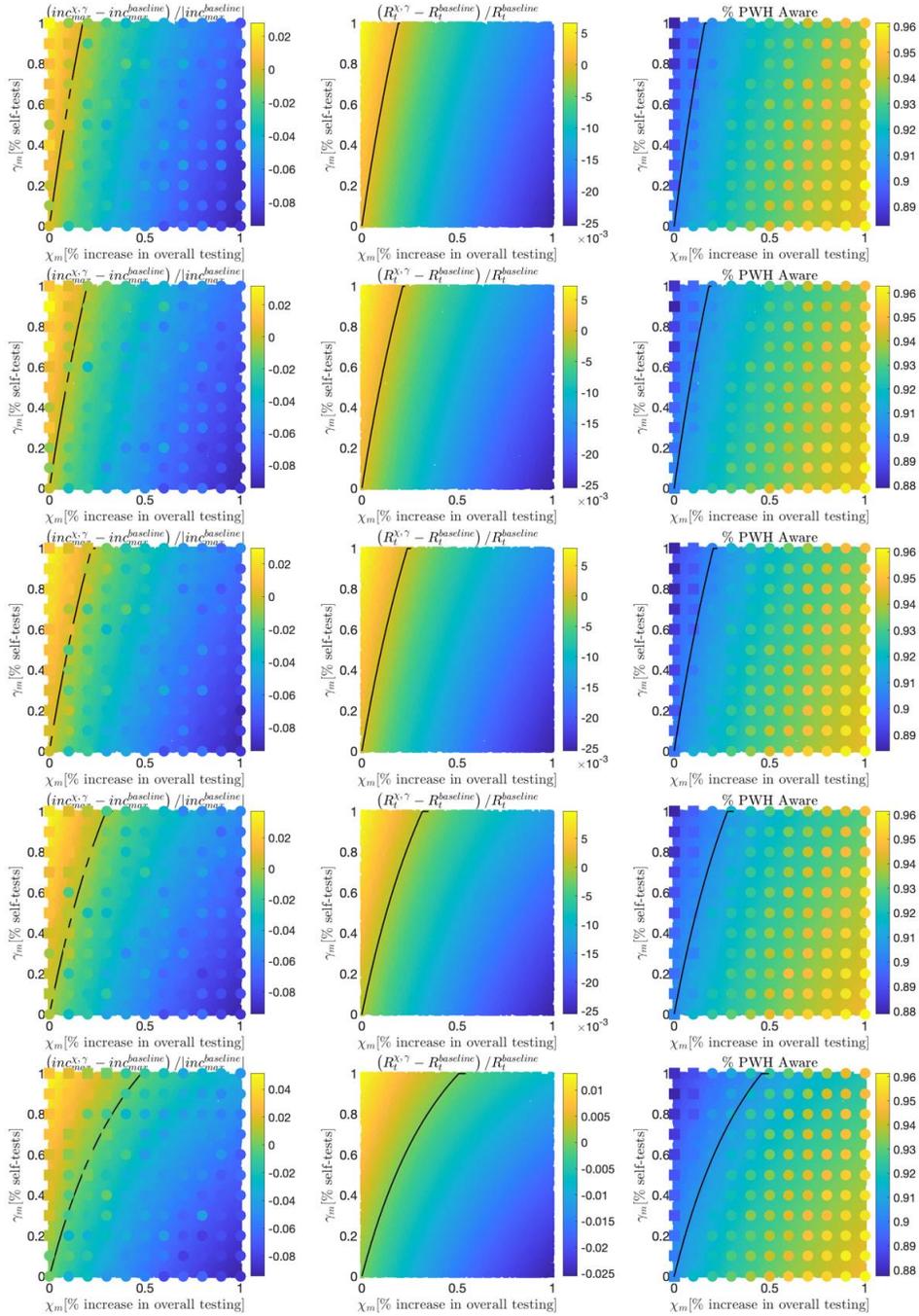